\documentclass[fonts]{icst}

\usepackage{moreverb}

\usepackage[breaklinks,colorlinks,bookmarksopen,bookmarksnumbered,linkcolor=ICSTblue,citecolor=ICSTblue,urlcolor=ICSTblue]{hyperref}
\usepackage{breakurl}
\usepackage{doi}
\usepackage{subcaption}
\usepackage{wrapfig}

\def\volumeyear{2014}

\newcommand{\crowdsensing}{crowdsensing }
\newcommand{\osensing}{opportunistic sensing }

\begin{document}

\runningheads{P.P Jayaraman, C. Perera, D. Georgakopoulos and A. Zaslavsky}{MOSDEN: A Scalable Mobile Collaborative Platform for Opportunistic Sensing Applications}

\title{MOSDEN: A Scalable Mobile Collaborative Platform for Opportunistic Sensing Applications}

\author{Prem Prakash Jayaraman\fnoteref{1}\affil{1}, Charith Perera\affil{1},
Dimitrios Georgakopoulos\affil{2} and Arkady Zaslavsky\fnoteref{2}\affil{1}}

\address{\affilnum{1} CSIRO Computational Informatics, Canberra, Australia 2601\\
\affilnum{2} School of Computer Science and Information Technology, RMIT University, GPO Box 2476, Melbourne VIC 3001
}

\abstract{Mobile smartphones along with embedded sensors have become an efficient enabler for various mobile applications including opportunistic sensing. The hi-tech advances in smartphones are opening up a world of possibilities. This paper proposes a mobile collaborative platform called MOSDEN that enables and supports \textit{opportunistic sensing} at run time. MOSDEN captures and shares sensor data across multiple apps, smartphones and users. MOSDEN supports the emerging trend of separating sensors from application-specific processing, storing and sharing. MOSDEN promotes reuse and re-purposing of sensor data hence reducing the efforts in developing novel \textit{opportunistic sensing} applications. MOSDEN has been implemented on Android-based smartphones and tablets. Experimental evaluations validate the scalability and energy efficiency of MOSDEN and its suitability towards real world applications. The results of evaluation and lessons learned are presented and discussed in this paper.}

\keywords{opportunistic sensing, crowdsensing, mobile middleware, mobile data analytics}

\fnotetext[1]{Corresponding author\\ 
Email: \email{\{prem.jayaraman, charith.perera, arkady.zaslavsky\}@csiro.au}\\
\email{dimitrios.georgakopoulos@rmit.edu.au}}
\fnotetext[2]{Prof. Zaslavsky is an International Adjunct Professor at StPetersburg National Research University of IT, Mechanics and Optics, Russia
}
\maketitle

\section{Introduction}

Today mobile phones have become a ubiquitous computing and communication device in people's lives \cite{Lane-Survey}. The mobile device market is growing at a frantic pace and it won't be long before it outnumbers the human population. It is predicted that mobile phones combined with tablets will exceed the human population by 2017 \cite{web1}. Current generation smartphones are equipped with plethora of features such as rich set of sensors (e.g. ambient light sensor, accelerometer, gyroscope, digital compass, GPS, microphone and camera) to enable on-the-move sensing and technologies such as NFC, Bluetooth, WiFi that enable them to communicate and interact with external sensors available in the environment. Smartphones have the potential to generate an unprecedented amount of data \cite{mobile-social-research} that can revolutionise many sectors of economy, including business, healthcare, social networks, environmental monitoring and transportation.

Mobile opportunistic sensing is one such new wave of innovative application popularly called collaborative community sensing or crowdsensing \cite{ganti, oppur-mobile}. Mobile Opportunistic sensing is an autonomous collaborative sensing approach that takes advantage of a population of users to measure large-scale phenomenon which cannot be measured using a user. Mobile opportunistic sensing applications require minimal user involvement (e.g. continuous computation of user activity passively i.e. in the background).  Opportunistic sensing applications \cite{mdm-paper} thrive on the widespread availability of smartphones and the diverse sets of data generated by these devices.  Date captured from an individual smartphone user can be used to infer the user's current context and activity. On the other hand, by fusing data from a multitude of smartphone user population, high level context information such as crowd activity within a given environment \cite{mdm-paper} can be inferred. In either form, the data generated by the smartphones is valuable and offers unique opportunities to develop novel and innovative applications. 

To date most efforts to develop opportunistic sensing/crowdsensing\footnote{In this paper, we use the terms \textit{opportunistic sensing } and \textit{crowdsensing} synonymously.} applications have focused on building monolithic mobile applications that are built for specific requirements \cite{pogo}. These application silos have very little capability for extensions and sharing of sensed data with a community of users often making the data available only within the application's context \cite{ganti}. However, to realise the greater vision of mobile collaborative \osensing we need a common extensible platform that facilitates easy development and deployment of collaborative \osensing applications on-demand. 

The key challenge here is to develop a platform that is autonomous, scalable, interoperable and supports efficient sensor data capturing, processing, storage and sharing. The autonomous ability of the system enables self-management and independent operations during device disconnections and off-line modes. We strongly believe that providing an easy-to-use, scalable platform to develop and deploy collaborative mobile \osensing applications will be significant. To this end, we propose a collaborative mobile sensing platform namely Mobile Sensor Data Engine (MOSDEN). MOSDEN is capable of functioning on multitude of resource-constrained devices (e.g. Raspberry Pi\footnote{http://www.raspberrypi.org/}) including smartphones. MOSDEN is a scalable platform that enables collaborative processing of sensor data. The MOSDEN platform follows component-based design philosophy allowing users/developers to implement custom data analytic algorithms (e.g. data mining algorithms \cite{minefleet}) and models to suit application requirements. Further, MOSDEN incorporates local processing, storage and sharing as a means to accomplish data reduction for Big Data applications. By limiting the continuous transmission of data to a centralised server which is typical of most mobile \osensing application, MOSDEN reduces bandwidth and power consumption. 
The key contributions of this paper are summarised as follows:

\begin{itemize}
\item We propose MOSDEN, a scalable, easy-to-use, interoperable platform that facilitates the development of collaborative mobile \osensing applications
\item We demonstrate the ease of development and deployment of a \osensing application using the MOSDEN platform
\item We experimentally evaluate and validate the scalability, performance and energy-efficiency of MOSDEN under varying collaborative \osensing workloads 
\end{itemize} 

The rest of the paper is organised as follows. Section II discusses related work. Section III considers a motivation scenario. Section IV presents the proposed MOSDEN platform architecture. Section V discusses MOSDEN implementation and Section VI presents MOSDEN platform evaluations and results. Section VI concludes the paper with indicators to future work.

\section{Related Work}

Numerous real and successful mobile \osensing applications have emerged in recent times such as WAYZ\footnote{\label{note1}http://www.wayz.com/} for real-time traffic/navigation information and Wazer2\footnote{https://www.wazer2.co.il/}  for real-time, location-based citizen journalism, context-aware open-mobile miner (CAROMM)\cite{mdm-paper} among others. Mobile \osensing applications \cite{ZMP003, ZMP008} thrive on the data obtained from heterogeneous sets of smart phones owned and operated by humans. Until recently mobile sensing application such as activity recognition (\textit{personal sensing)}, where people's activity (e.g. walking, talking, sitting) is classified and monitored, required specialised mobile devices \cite{activity-recognition-mobile-sensing, thesis-wearable}. This has significantly changed with advent of smartphones equipped with on-board sensing capabilities. More recently, research efforts have focused on development of activity recognition, context-aware \cite{ZMP007} and data mining models for smartphones \cite{activty-recognition-gomes, minefleet, contextphone} that leverage on smartphone's processing and on-board sensing capabilities.

Recent efforts to build \osensing application have focused on building monolithic mobile applications that are built with specific purpose and requirements. The MetroSense \cite{metrosense} project at Dartmouth is an example of one such \osensing system. The project aims in developing classification techniques, privacy approaches and sensing paradigms for mobile phones. The CenceMe \cite{Miluzzo:2007:CIS:1775377.1775379} project under the MetroSense umbrella is a personal sensing system that enable members of social networks to share their presence. The CenceMe application incorporates mobile analytics by capturing user activity (e.g., sitting, walking, meeting friends), disposition (e.g., happy, sad, doing OK), habits (e.g., at the gym, coffee shop today, at work) and surroundings (e.g., noisy, hot, bright, high ozone) to determine presence. The CenceMe system comprises two parts, the phone software and back-end software. The phone software is implemented on a Nokia N95 running Symbian operating system. The phone software is developed in Java Micro Edition (JME) which interfaces with Symbian C++ modules controlling the hardware.

MineFleet \cite{minefleet} is a distributed vehicle performance data mining system designed for commercial fleets.  In MineFleet, dedicated patented custom built hardware devices are used on fleet trucks to continuously process data generated by the truck. MineFleet system comprises an on-board data stream mining module that performs extensive processing of data using various statistical and data stream mining algorithms.  This data stored locally is transmitted to an external MineFleet Server for further processing when network connectivity is available. 

Crowdsourcing data analytics system (CDAS) \cite{CDAS} is a crowd sourcing framework in which, the participants are part of a distributed crowdsourced system. The CDAS system enables deployment of crowdsourcing applications that require human involvement for simple verification tasks using Amazon Mechanical Turk (AMT). E.g. users in the system would be sent a picture for identification by a centralised task distributor. Humans users participating in the crowdsourcing application respond with appropriate answers. The results from human workers are combined to compute the final result. The CDAS system incorporates complex analytics that enables it to disseminate jobs, obtain results and compare results obtained from different workers to determine the correct one. GeoCrowd \cite{Kazemi:2012} is another crowdsourcing system that employs spatial characteristics to estimate task assignments among user populations. Mobile edge capture and analysis middleware for social sensing applications (MECA) \cite{ye_meca:_2012} is another middleware for efficient data collection from mobile devices in a efficient, flexible and scalable manner. MECA provides a platform by which different applications can use data generated from diverse mobile data sources (sensors). The proposed MECA architecture has three layers comprising data layer (mobile data sources – mobile phones), edge layer (base stations that select and instruct a device or group of devices to collect data and process data), phenomena/application layer (the back-end that determines the edge nodes to process application request).

Applications like Waze\footnote{\label{note1}http://www.wayz.com/}, MetroSense \cite{metrosense} and MineFleet \cite{minefleet} are built around specific data handling models (e.g. GPS for Waze, Microphone for MetroSense and Data mining algorithms for object monitoring) and application requirements reducing its re-usability. On the other hand, frameworks like CDAS\cite{CDAS}, GeoCrowd \cite{Kazemi:2012} and MECA \cite{ye_meca:_2012} offload processing to centralised servers increasing bandwidth usage and making them less suitable for working in off-line modes. Moreover, in these frameworks, the smartphones are viewed as mere data collection or user response collection devices lacking capabilities to implement localised data analytics. On contrast, the proposed MOSDEN platform is developed with the design goal of 1) re-usability, 2) ease of use, 3) ease of development/deployment, 4) scalability, 5) easy interface to access both on-board and external sensors, 6) support for on-board complex data analytics and 7) distributed mobile data sharing. The MOSDEN platform provides the application developer with implementation options i.e. choice of using processing on the smartphone and/or processing at the server. It also provides extensions to implement mobile distributed load-balancing that can determine on-demand the best location to process data. We note, discussions on load-balancing and task allocation to different collaborative MOSDEN smartphone devices are outside the scope of this paper. The MOSDEN platform promotes a distributed collaborative sensing infrastructure where each MOSDEN instance running on a smartphone is self-managed. In our previous work \cite{collabo-prem} we proposed the architecture of MOSDEN supported by evaluations focusing on system performance. In this paper, we further analyse and present evaluation results of MOSDEN's scalability performance and energy-efficiency under varying workloads typical of collaborative \osensing applications. Specifically, we have critically evaluated the energy efficiency of MOSDEN platform when performing operations like continuous sensing and sensing/sending. We also evaluate MOSDEN's query processing efficiency when answering to distributed queries from multiple MOSDEN instances in a collaborative experimental setup.

\section{\label{section3} Opportunistic Sensing Scenario}

\begin{figure*}[t!]
\centering
\includegraphics[scale=0.3]{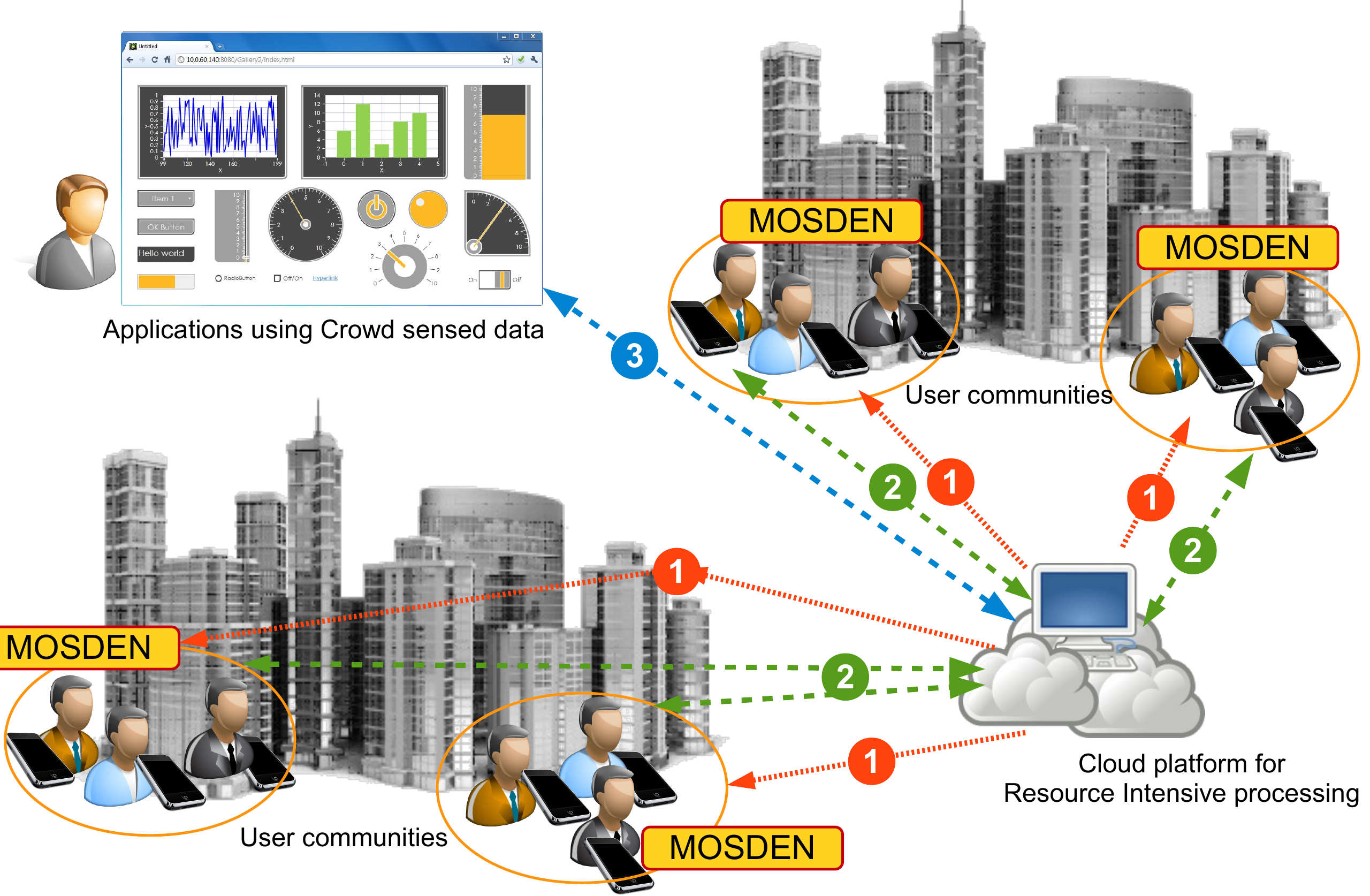}%
\captionsetup{justification=centering, labelsep=period}
\caption{Environmental Monitoring - Mobile Opportunistic Sensing Scenario}
\label{scenario}
\end{figure*}

In this section we present a motivating scenario that explains the need for a scalable, collaborative, mobile sensing platform like MOSDEN. The scenario under consideration is an environmental monitoring application (e.g. noise pollution) in smart cities as depicted in Fig. \ref{scenario}. 

In step (1), a remote-server (cloud-based) registers the interest for data within user communities. In the example depicted in Fig. \ref{scenario}, the user communities are grouped based on location. In step (2), the data captured and processed on the smartphones are made available to the remote-server. In step (3), the opportunistic sensing application obtains data from the remote-server for further processing and visualisation. The above scenario is a typical case for many \osensing applications that require data from diverse user communities. The same approach is applied to another \osensing application that computes air pollution within the environment. To accomplish this requirement, the smartphone will also have to rely on external sensors that are part of a smart city infrastructure to obtain air pollution data. 

Using a monolithic approach may results in developing a niche class of applications built for a single purpose. Such an application may not be scalable/adaptable to work in other scenarios which is a major obstacle. E.g. the algorithm required to process noise data is different to air pollution computation. Moreover, such an \osensing application is hard-wired making it extremely difficult to make changes to different parts of the code. E.g. adding new interface to communicate with external sensors to collect air pollution data.

To achieve the level of extensibility and scalability to support a range of different \osensing application requirements, the \osensing platform needs to follow the design principle of separating the sensor capturing operation and application specific processing. This will in turn promote on-demand application composition. Further, the platform needs to support real-time data collection, processing and storage, ability to incorporate application specific data analytic algorithms/models, energy-efficient operation and autonomous functions i.e. ability to work with minimal user interaction and with support for off-line modes. The proposed MOSDEN platform supports the above mentioned features natively.

\section {MObile Sensor Data ENgine (MOSDEN): A collaborative mobile \osensing platform}
\label{section4}

We propose MOSDEN, a \crowdsensing platform built around the following design principles:
\begin{itemize}
\item Separation of data collection, processing and storage to application specific logic
\item A distributed collaborative \crowdsensing application deployment with relative ease
\item Support for autonomous functioning i.e. ability to self-manage as a part of the distributed architecture
\item A component-based system that supports access to internal and external sensor and implementation of domain specific models and algorithms
\end{itemize}

These design principles address the obstacles mentioned in Section \ref{section3}. The proposed MOSDEN platform overcomes the key barriers of developing and deploying scalable collaborative mobile \osensing applications.

\begin{figure*}[t!]
 \centering
 \includegraphics[scale=0.35]{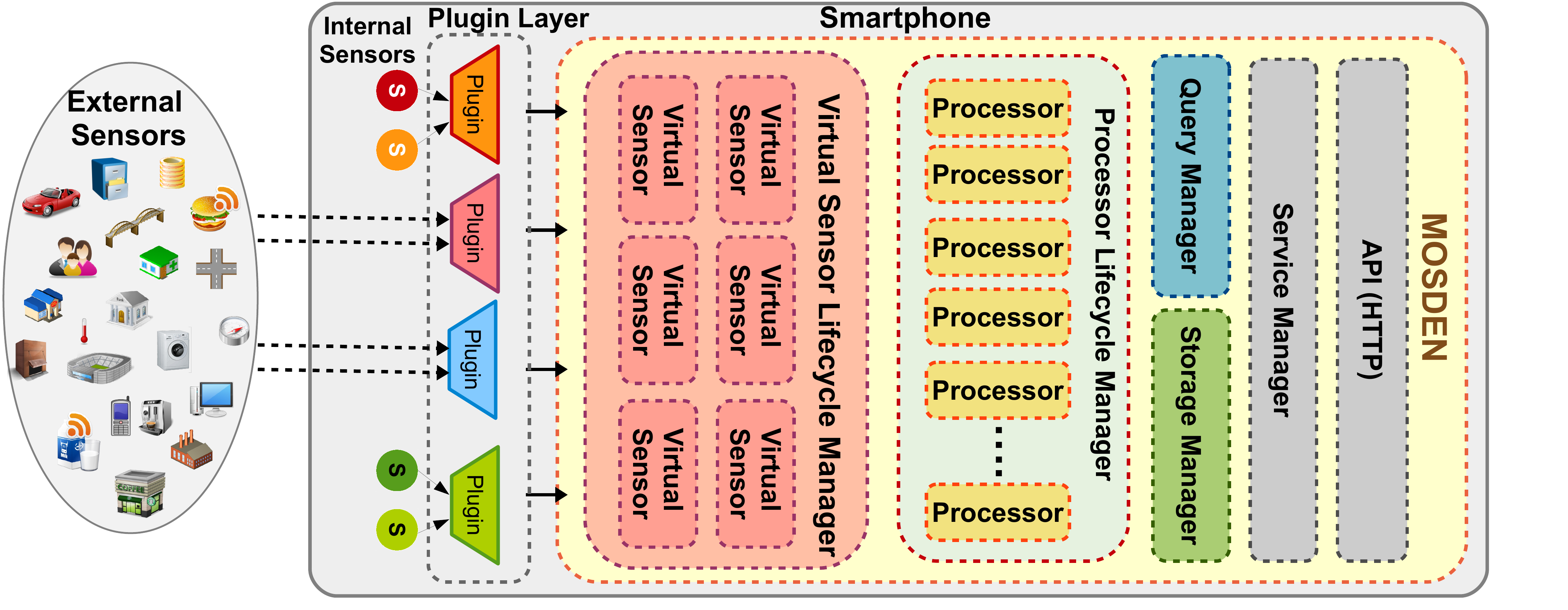}
\captionsetup{justification=centering, labelsep=period}
 \caption{MOSDEN Platform Architecture}
 \label{sys-arch}	
\end{figure*}

\subsubsection{Architecture}

MOSDEN platform follows the design principle of Global Sensor Network (GSN) \cite{P227}. GSN is a sensor network middleware developed to run on high-powered computing devices (e.g. servers and cloud resources). GSN presents a unified middleware approach that facilitates acquisition, processing and storage of sensor data. We reuse the concept of virtual sensors proposed by GSN. A virtual sensor is a abstraction of the underlying data source (e.g. wireless sensor network). Since, GSN was not developed for resource constrained environment, we made significant enhancement to GSN when designing and implementing MOSDEN described in the following section.  MOSDEN follows a component-based architecture allowing extensibility without modifying the existing codebase. The architecture of the proposed MOSDEN platform is presented in Fig. \ref{sys-arch} followed by description of each component. 

\begin{figure}[t!]
 \centering
 \includegraphics[scale=0.35]{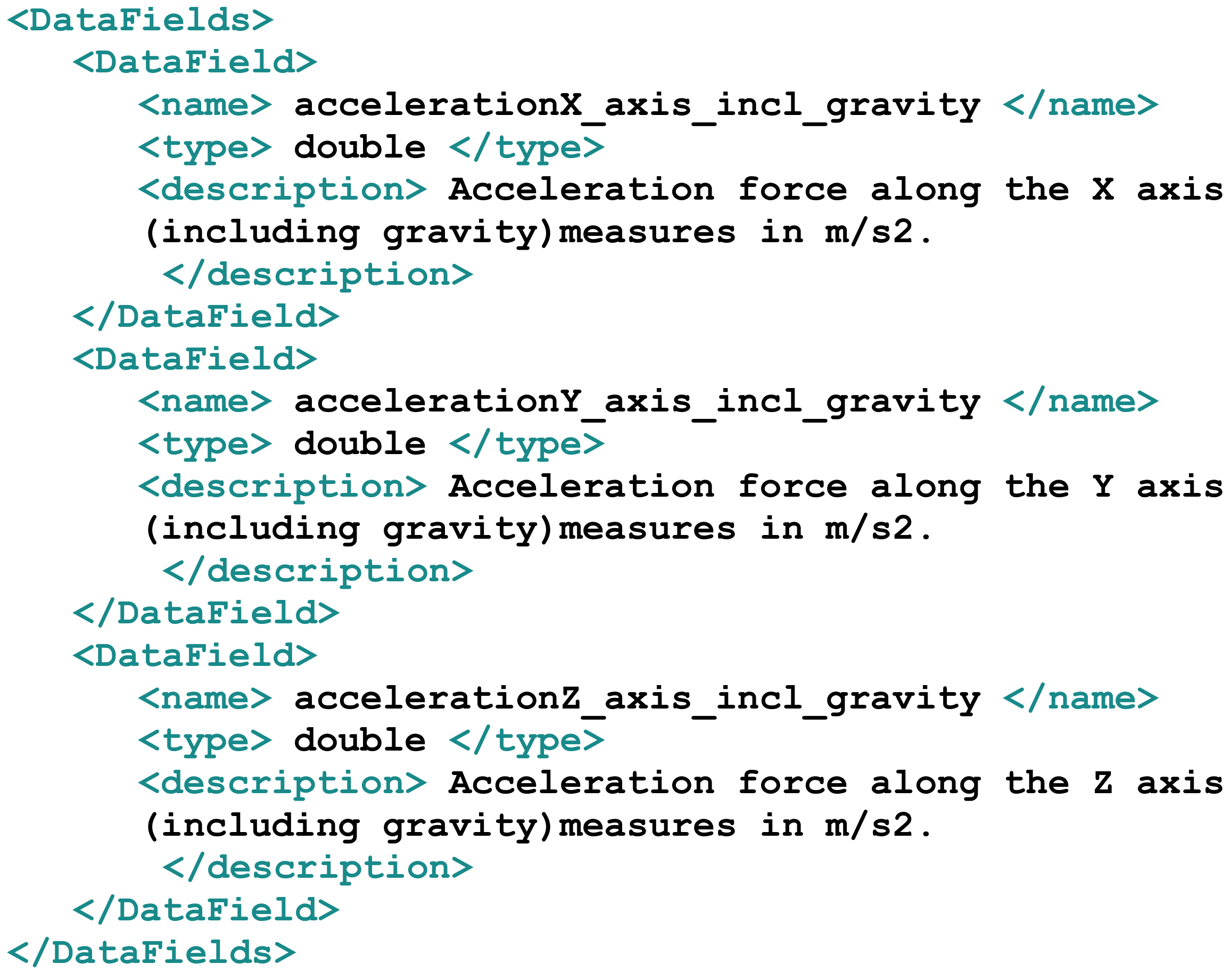}
\captionsetup{justification=centering, labelsep=period}
 \caption{A Conceptual Description of MOSDEN Plugin}
 \label{plugin}	
\end{figure}

\begin{description}
\item[\textit{Plugin:}] 
In MOSDEN, we introduce the concept of Plugins. In GSN, a developer had to implement wrappers to accommodate new sensor data sources into the system. To accommodate a new wrapper, the system had to be recompiled and redeployed. This approach is not very practical especially for real-time applications. Hence, we introduced a plugin-based approach to overcomes this challenge. The Plugins are independent applications that communicates with MOSDEN. MOSDEN uses a discovery mechanism specific to the implementation platform to discover the list of plugins installed in the system. E.g. in case of android operating system, a plugin discovery services provides a list of registered plugins and their description. The key function of the plugin is to describe how to interface with a sensor that needs to be connected with MOSDEN. We have developed a plugin descriptor that \osensing application developer can use to implement plugins to interface with new sensors. A conceptual description of the plugin is shown in XML format in Fig. \ref{plugin}. 

\item[\textit{Virtual Sensor:}]	The virtual sensor is an abstraction of the underlying data source from which data is obtained. The virtual sensor can perform stream level operation e.g. fuse data from multiple sensor sources (on-board, external sensors connected to the mobile device and remote MOSDEN instances). The virtual sensor can be used to specify configurations to capture data from distributed MOSDEN instances. In such situations, the query and service manager at the remote MOSDEN instance is responsible for processing the query. The virtual sensor file dynamically links the sensor plugins with MOSDEN platform via an XML configuration file. The virtual sensor lifecycle manager is responsible to manage the instantiation, updation and removal of virtual sensor resources. 

\item[\textit{Processors:}]	The processor classes are used to implement application specific data analytics models and algorithms. For example, a Fast Fourier Transform (FFT) algorithm to compute the decibel level from microphone recordings or a data mining algorithm to perform high speed data stream clustering and classification \cite{minefleet}.

\item[\textit{Storage Manager:}] The function of the storage manage is to store the data acquired from the virtual sensor and processor workflow. The storage manager uses a data collection window to delete old data. This function of the data collection windows is very similar to sliding window protocol. The virtual sensor configuration file is used to configure the data collection window size during application deployment. The window size can be modified during runtime.

\item[\textit{Query Manager:}]	The query manager is responsible to resolve and answer queries from external entities. An external entity can be another MOSDEN instance, a user or an application querying for data collected by the smartphone. The query manager employs a queuing mechanism to resolve incoming queries. The local storage and query processing functionality of MOSDEN is a key enabler of off-line mode operations.

\item [\textit{Service Manager:}]	The service manager is responsible to manage subscriptions to data from external entities. The service manager registers subscription request and depending on the mode of data delivery (e.g. persistent/non-persistent) will deliver available data to the requested external source when possible. The service manager is responsible to handle data connections with external data sources. The service manager is specifically designed to manage the working of MOSDEN in resource constrained environments where frequent disconnection occurs.

\item[\textit{API Manager:}]	The application programmable interfaces (APIs) provides a standard way to subscribe and access data to/from MOSDEN instances. The API requests are received and processed over HTTP. Request received via the API are passed to the service manager for further processing and management.

\end{description}

Each MOSDEN instance running on the mobile smartphones can run with minimal user interaction. It can received and register a data capture request/processing from a remote-server (e.g. cloud-based). MOSDEN then works in the background processing the request by collecting, processing and storing the requested data locally. When the processed data is required by the application running at the remote-server, it can query the specific MOSDEN instance running on user smartphone for the data. MOSDEN realises a true distributed collaborative system architecture as it has the ability to function independent of the remote-server.

As depicted in the architecture, each individual MOSDEN instance is self contained and managed and is capable of working in mobile environments that encounter frequent disconnections. The use of APIs to communicate between instances encourages collaborative workload sharing and processing. The plugin based approach increases re-usability and promotes interoperability allowing MOSDEN to communicate with any sensors both internal and external.  This remove the burden on \osensing developer. Further, the use of a component-based architecture and a workflow style processing allows system developers to implement and integrate domain specific data analytics algorithms with relative ease.  Moreover, the MOSDEN platform enables the development of mobile \osensing applications that can scale from an individual user to a community. E.g. an individual personal fitness monitoring application to a group activity recognition application involving a community of users can be developed and deployed using the MOSDEN platform.

\section{Implementing a Collaborative \osensing Application using MOSDEN}
%For example, the platform can be used to develop a individual personal fitness monitor application taking advantage of on-board sensing capabilities to group activity recognition application that computes activities of a group by obtaining activity information from a community of users.

In Section \ref{section3} we presented an environmental monitoring scenario to determine the noise pollution level using the data obtained from a community of user. Using the information obtained from the user communities, a \osensing application running on a remote-server (e.g. cloud) can further analyse and visualise the noise pollution level of a smart city. Each user community in this scenario is grouped by their locations.

\begin{figure}[b!]
 \centering
 \includegraphics[scale=0.35]{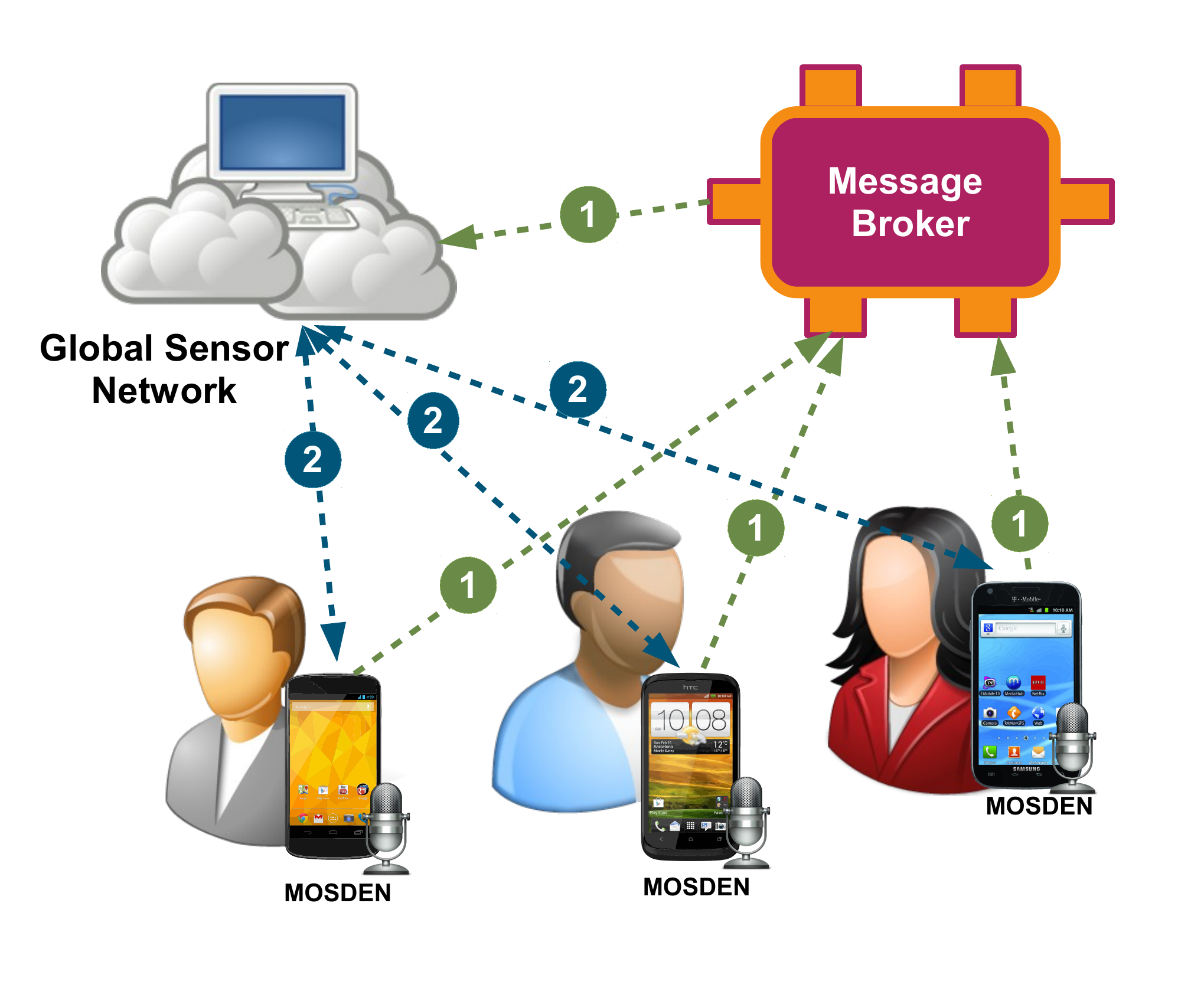}
\captionsetup{justification=centering, labelsep=period}
 \caption{Implementation of Opportunistic Sensing Application using MOSDEN}
 \label{Implementation of Opportunistic Sensing Application using MOSDEN}	
\end{figure}

In this section we present a detailed description of the noise pollution \osensing proof-of-concept application implemented using MOSDEN platform. Fig. \ref{Implementation of Opportunistic Sensing Application using MOSDEN} presents the overview of the application implemented on MOSDEN platform. In the scenario depicted in \ref{Implementation of Opportunistic Sensing Application using MOSDEN}, in step (1) MOSDEN instances running on the smartphone registers with the cloud GSN instance (clients are aware of server's IP address during configuration process). Once registration is complete in step (2) the cloud GSN instance registers its interest to receive noise data from MOSDEN running on the smartphones. When data is available, MOSDEN on the smartphones stream the data to the cloud GSN. The data transfer process could employ push or pull techniques over a persistent or non-persistent connection depending on application requirement. In this specific example we implemented a pull-based approach where GSN running on the remote-server will specifically query each MOSDEN instance running on the smartphones.  

 \begin{figure*}[!t]
  \centering
	 \includegraphics[scale=0.9]{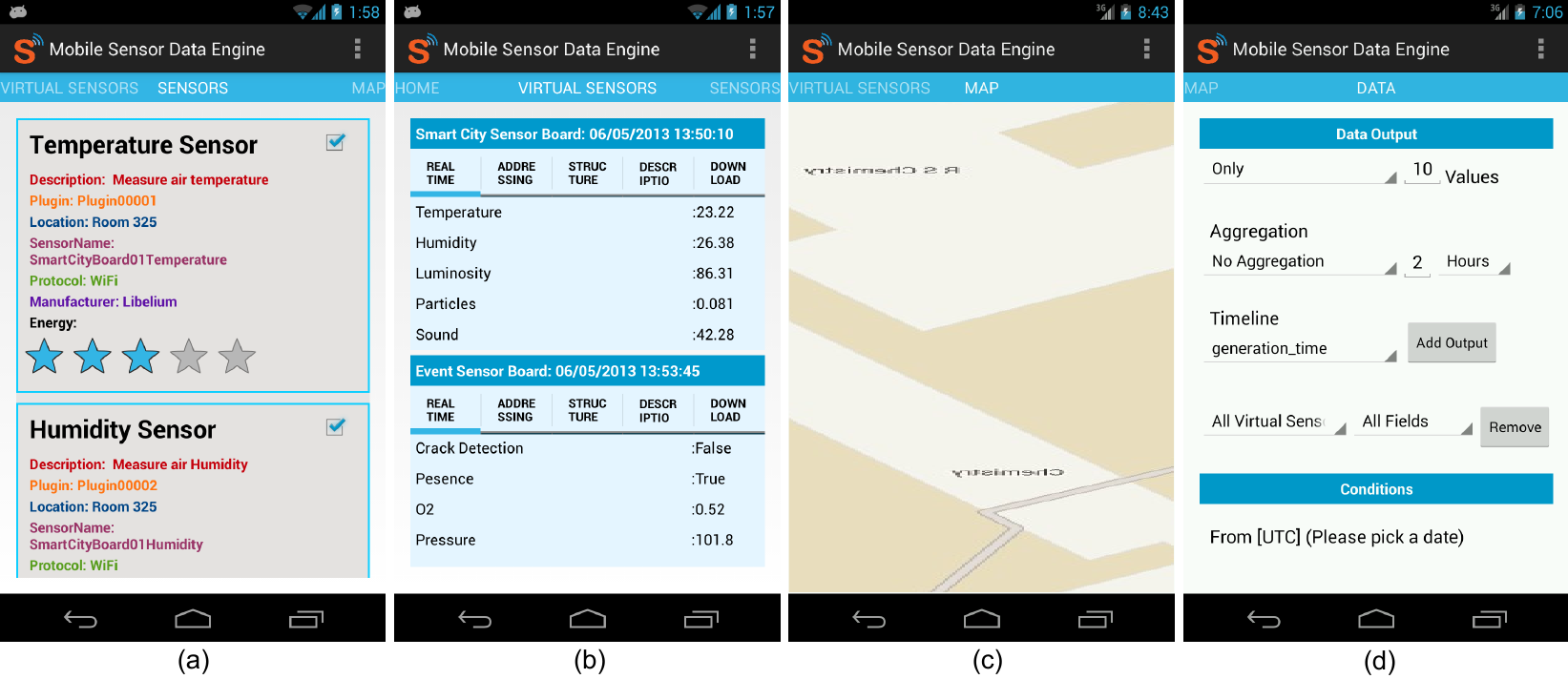}
	\captionsetup{justification=centering, labelsep=period}
	 \caption{MOSDEN screenshots: (a) List of sensors connected the MOSDEN; (b) List of  virtual sensors currently running on the MOSDEN and their details; (c) Map that shows sensor locations; and (d) Interface for data fusing and filtering}
	 \label{MOSDEN}	
 \end{figure*}

The MOSDEN reference architecture presented in Fig. \ref{sys-arch} has been implemented using the Android\footnote{http://www.android.com/} SDK platform. We deployed the noise pollution application developed on the MOSDEN platform on a set of smartphone and tablet devices that simulate a user community. To compute the noise decibel level, we implemented a modified version of the processing class from Audalyzer open source project\footnote{https://code.google.com/p/moonblink/}. The microphone sensor on the smartphones was used to obtain raw sound recordings. Code to interface with the microphone sensor was already available as a part of the MOSDEN platform via plugins (we have developed plugins for on-board sensors and few selected external sensors e.g. waspmotes - http://www.libelium.com/products/waspmote/). For our proof-of-concept implementation, we implemented GSN in the cloud that queries data from individual MOSDEN instances running on user mobile devices. A MOSDEN instance once deployed on the smartphone/tablet registers itself with the GSN in the cloud. As we stated earlier, the design of MOSDEN makes it easily extensible to suit any \osensing application requirements. To validate this, we implemented the registration process via a message broker as depicted in Fig. \ref{Implementation of Opportunistic Sensing Application using MOSDEN}. Along with the registration, each MOSDEN instances also updates the cloud GSN instance with a list of available sensors. We note, MOSDEN API extends support to any form of registration. It is the responsibility of the \osensing application developer to choose the most appropriate registration process best suited to the application.

\begin{figure}[h]
 \centering
 
 \begin{subfigure}[h]{0.3\textwidth}
  \centering
	 \includegraphics[width=\textwidth]{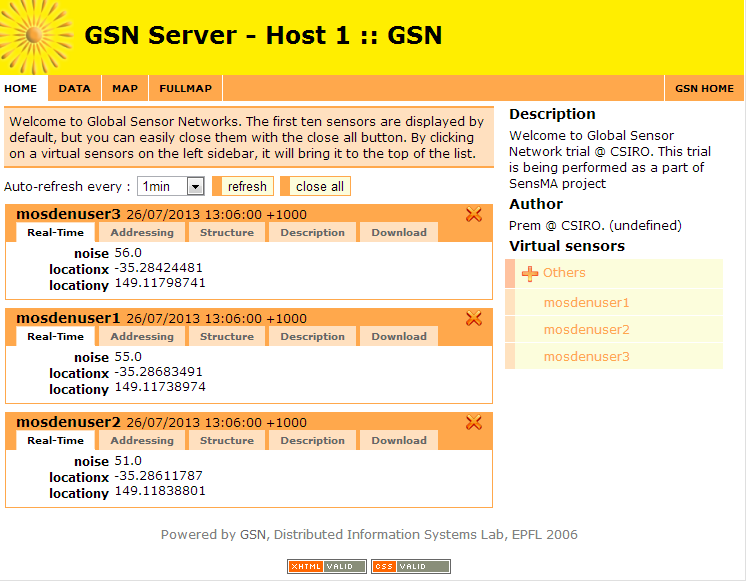}
	\captionsetup{justification=centering, labelsep=period}	
	 \caption{GSN Sensor Registration Screenshot}
	 \label{GSN-screenshots1}	
 \end{subfigure}

 \begin{subfigure}[h]{0.3\textwidth}
  \centering
	 \includegraphics[width=\textwidth]{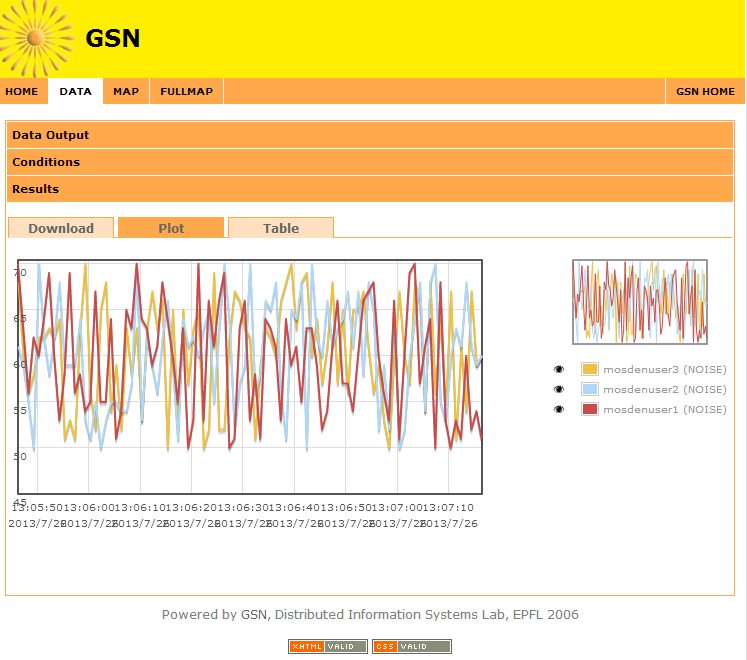}
	\captionsetup{justification=centering, labelsep=period}	
	 \caption{GSN Noise Plot Screenshot}
	 \label{GSN-screenshots2}	
 \end{subfigure}
 \captionsetup{justification=centering, labelsep=period}
 \caption{Opportunistic Sensing Noise Pollution Application Screenshots}
 \label{oppsensing-app}
\end{figure}

MOSDEN's API enables it to query data from any other MOSDEN instances. Hence, it is to be noted that the GSN running in the cloud instance could be replaced by another smartphone running MOSDEN. In such a scenario, the MOSDEN instance is also responsible to query data from other smartphones, performs further data analytics on collected data and perform visualisation. The analytics and visual components required to accomplish the previously states functions can be easily integrated with MOSDEN as individual components. Screenshots of the MOSDEN implementation on Android smartphone and GSN in the cloud are illustrated in Fig. \ref{MOSDEN} and \ref{oppsensing-app}. We note, the default version of GSN with no enhancements was used to demonstrated the proof-of-concept implementation. Fig. \ref{GSN-screenshots2} depicts the noise graph computed from 3 MOSDEN users. In this example, we have plotted the noise data individually. To demonstrate the ease of deploying complex \osensing applications using MOSDEN, Table \ref{LOC-sensing} presents the total lines of code that was required to develop the previously mentioned noise pollution monitoring application. We note, the configuration files required by MOSDEN to implement a new application are Virtual Sensor and Plugin Wrapper. The plugin application to capture microphone data and data analytics component are application specific implementations external to MOSDEN code and would change depending on application complexity. In the current implementation, the plugin application is a standard android sensor service implementation.

\begin{table}[t]
\caption{LOC to develop a Noise Pollution Monitoring Application}
\label{LOC-sensing}
\centering
\begin{tabular}{| p{5.5cm} | c |}
\hline
Component & Lines of Code\\
\hline
Virtual Sensor (MOSDEN) & 30 Lines\\
\hline
Plugin Wrapper (MOSDEN) & 190 Lines\\
\hline
Plugin Application: Capture Microphone data (External) & 75 lines \\
\hline
Data Analytics: Decibel Computation FFT (External) & 194 Lines \\
\hline
\end{tabular}

\end{table}

\subsubsection{Benefits of MOSDEN Design}

The proposed MOSDEN model is architected to support scalable, efficient data sharing and collaboration between multiple application and users while reducing the burden on application developers and end users. The scalable architecture can easily be orchestrated for \osensing applications that range from an individual to a community of users. It facilitates easy sharing of data among large community of users which is a vital requirement for \osensing applications.

By separating the data collection, storage and sharing from domain-specific application logic, our platform allows developers to focus on application development rather than understanding the complexities of the underlying mobile platform. In fact, our model hides the complexities involved in accessing, processing, storing and sharing the sensor data on mobile devices by providing standardised interfaces that makes the platform reusable and easy to develop new application. This we believe will is critical and will significantly reduce the time to develop new innovative \osensing applications. Since, MOSDEN is designed as a component-based architecture, it provides easy interfaces to implement application specific data analytic models and algorithms.

Further, our model works in the background with minimal user interaction reducing the burden on smartphone users. By providing support for processing and storage on the device, we also reduce frequent transmission to a centralised server as compared to current \osensing frameworks. The potential reduction in data transmission has the following benefits: 1) saves energy on users' mobile device; 2) reduces network load by avoiding long-running data transmissions and 3) reduces data transmission costs by limiting continuous data transmissions.

%This results in the following benefits: 1) Better access to high quality real-time data, 2) 
The benefits of MOSDEN Design can also be articulated from a Big Data perspective. Evidence from Big Data applications like Phenonet\cite{phenonet} indicate, only 0.1\% of the data collected for scientific plant experiments (from 1 million data points) represent golden data points. A typical Big Data approach introduces the challenges of capturing, storing filtering and analysing such volumes of data streams in remote cloud servers. Such an approach is both time and resourcing consuming. An alternative approach that we propose is to leverage on MOSDEN-like architecture leveraging on distributed local data analytics, storage, retrieval on-demand and off-line functioning capability for the following benefits: 1) Data reduction: data filtering near the data capturing location using local data analytics e.g. using statistical approaches to filter unwanted and erroneous data; 2) Relevant Data stream Selection: ability to selectively choose relevant data sources and query these sources for data depending on application needs. This will again reduce transfer and processing of large amounts of data to a remote-location; 3) Better real-time access to data on-demand and 4) Reduction in resource and bandwidth consumption due to collaborative distributed data analytics, storage and querying.

To validate the performance of the proposed MOSDEN platform to support scalable,energy and resource efficient data sharing and collaboration, in the next section we present MOSDEN platform evaluations. To this end, we evaluate MOSDEN under extreme loads typical of collaboratively \osensing application scenarios.

\section{Evaluation of MOSDEN Platform}
In this section, we present the details of our experimental test-beds and evaluation methodology. Further, we discuss the results and present the lessons learnt from  experimental evaluations. The overall objective of the experimental evaluations is to examine the scalability, resource consumption, performance and energy consumption of MOSDEN platform in collaborative environments. Scalability of MOSDEN under collaborative application scenarios is tested by experimenting MOSDEN's ability to handle growing amount of sensing and querying tasks in a capable manner.

%More specifically, we evaluate MOSDEN capability to capture data from on-board and external sensors  

\subsection{Experimentation Testbed and Setup}
\label{sec:E:Experimentation Testded}		

For the evaluation of the proof of concept implementations, we used four mobile devices and a laptop. From here onwards we refer them as D1, D2, D3, D4, and D5 respectively. The technical specifications of the devices are as follows.

\begin{itemize}
\item \textbf{Devices (D1 \& D4):} Google Nexus 4 mobile phone, Qualcomm Snapdragon S4 Pro CPU, 2 GB RAM, 16GB storage, Android 4.2.2 (Jelly Bean) 

\item \textbf{Devices (D2 \& D3):} Google Nexus 7 tablet, NVIDIA Tegra 3 quad-core processor, 1 GB RAM, 16GB storage, Android 4.2.2 (Jelly Bean)

\item \textbf{Device (D5):} ASUS Ultrabook Intel(R) Core i5-2557M 1.70GHz CPU and 4GB RAM (Windows 7 operating system)
\end{itemize} 

\begin{figure}[t!]
 \centering
 \includegraphics[scale=0.35]{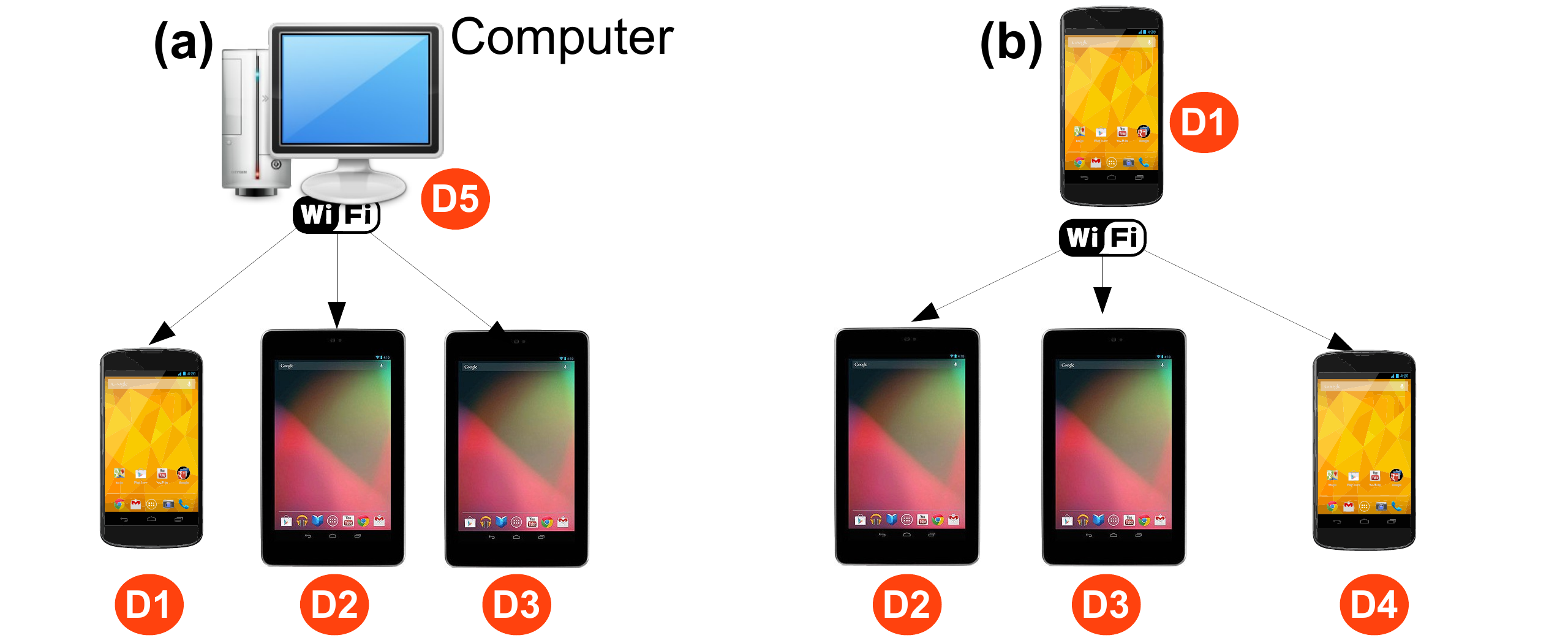}
\vspace{-0.22cm}
\captionsetup{justification=centering, labelsep=period}	
 \caption{Experimental Testbed has been configured in two different ways: (a) Setup 1: Three mobile devices are connected to a laptop and 
(b)  Setup 2: three mobile devices are connected to another  mobile device.}
 \label{Figure:Setup}	
\vspace{-0.43cm}	
\end{figure}

For experimentation, we devised two setups as illustrated in Fig. \ref{Figure:Setup} and evaluated the proposed framework in each setup independently. For ease of illustration, in each setup, the parent node performs the operations of the remote-server including issuing queries to fetch data and process and store obtained data while child nodes perform client operations including data capturing, local analytics, storage and query processing. The proposed MOSDEN platform can be deployed in either roles i.e. remote-server or client depending on application requirements. In our setup depicted in Fig. \ref{Figure:Setup}a, the MOSDEN platform on the mobile device (D1, D2, D3) assumes the role client while in Fig. \ref{Figure:Setup}b, the MOSDEN platform on the mobile device (D1, D2, D3, D4) assumes the role of both remote-server and client. The laptop computer (D5) in Fig. \ref{Figure:Setup}b is configured to run GSN engine \cite{P227}. The MOSDEN architecture promotes a distributed collaborative system with connection between MOSDEN instances (client and server) maintained and managed independently as peers. The term "client" and "server" is used to specify the temporary role of the mobile devices during experiments i.e. server is responsible to answer to user queries while clients only collect data. For example, in setup 2, the mobile device (D1) running MOSDEN can also perform local sensing and respond to requests from other MOSDEN instances transforming this setup into a hierarchical peer-to-peer architecture. At any time using minimal configurations, MOSDEN on mobile devices can perform the role of both client and server making the collection of MOSDEN's devices peers. We aim to extend out experiments in future for peer-to-peer scenarios. It is to be noted, the use of client server terminology does not limit MOSDEN platform to only client server setup, rather is used for ease of illustration. Further, for evaluations purposes, we chose a setup with 4 devices in different configurations. Experimental evaluations show that MOSDEN can scale to n number of devices when working in either client or server modes.

%In this section, we explain the experiment setup and objectives behind each experiment we conducted in detail. Next section discusses the results and lessons learnt in detail. 
The maximum number of sensors was set to 13 and kept fixed throughout the experiments\footnote{All the sensors available on the given device has been used (e.g. In D1: accelerometer, microphone, light, orientation, proximity, gyroscope, magnetic, pressure).}. In all the evaluations, CPU usage (consumption) is measured in units of jiffies\footnote{In computing, a jiffy is the duration of one tick of the system timer interrupt. It is not an absolute time interval unit, since its duration depends on the clock interrupt frequency of the particular hardware platform}. Sampling rate for all the sensors connected to MOSDEN is one second. 

In Fig. \ref{Figure:Setup}(a) and (b), a query in the form of a \textit{request} is sent from the server to MOSDEN client instances. Depending on the number of sensors queried on MOSDEN instances, the number of requests increase. We use the term \textit{'MOSDEN client'} to refer to mobile devices where MOSDEN act as a client such as D1, D2 and D3 in setup 1 in Fig. \ref{Figure:Setup}(a) and D2, D3 and D4 in setup 2 in Fig. \ref{Figure:Setup}(b)). We use the term  \textit{'MOSDEN server'} to refer to a mobile device where MOSDEN performs the role of a  server such as D1 in setup 2 in Fig. \ref{Figure:Setup}(b)).

\subsection{Experimental Results and Discussion}
In this section, we will present discussions of each experiment we conducted in detail.
\subsubsection{CPU and Memory Consumption Experiment}
In this experiment, we evaluate the CPU and memory usage of MOSDEN platform functioning as client and server in setups (a) and (b) illustrated in Fig. \ref{Figure:Setup}. To experimentally evaluate MOSDEN client's resource consumption, we conducted two experiments. In the first experiments, we computed the total CPU and memory consumption when performing sensing (on-board sensors), processing and local storage (henceforth we use the term \textit{sensing} to represent the 3 operations). In the second experiment, we also include resource consumption incurred due to data transmission over Wi-Fi For MOSDEN server (D1 in Fig. \ref{Figure:Setup})(b)), the experiment only considered the resources consumed to process queries from distributed MOSDEN client instances (D1, D2 and D3 in Fig. \ref{Figure:Setup})(a)). 

\begin{figure}[b!]
 \centering
 \includegraphics[scale=0.45]{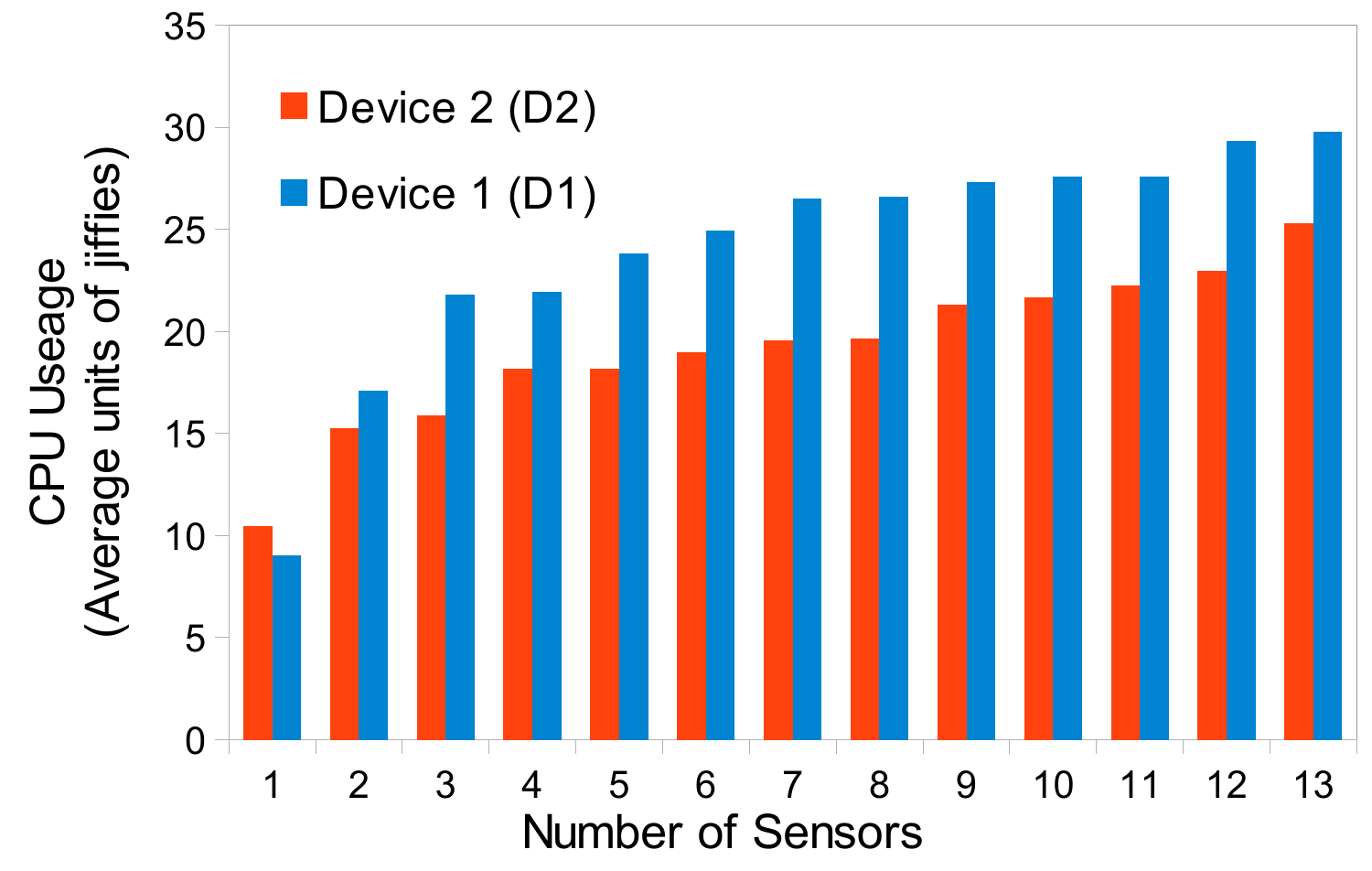}
\captionsetup{justification=centering, labelsep=period}	
 \caption{Comparison of CPU Usage by MOSDEN Client - Sensing}
 \label{Experiment1a}		
\end{figure}

\begin{figure}[t!]
 \centering
 \includegraphics[scale=0.45]{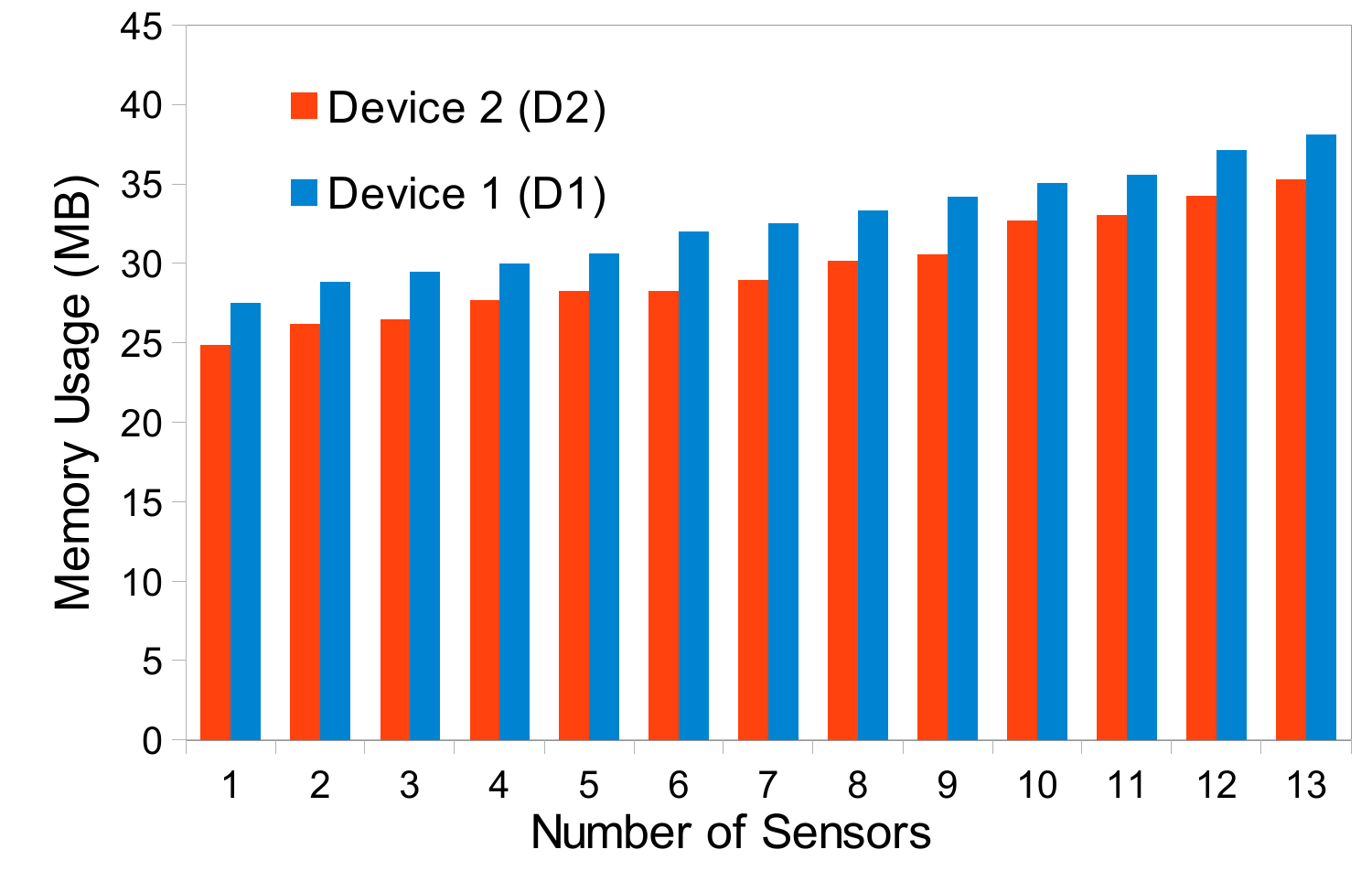}
\captionsetup{justification=centering, labelsep=period}	
 \caption{Comparison of Memory Usage by MOSDEN Client - Sensing}
 \label{Experiment2a}		
\end{figure}

\begin{figure}[t!]
 \centering
 \includegraphics[scale=0.45]{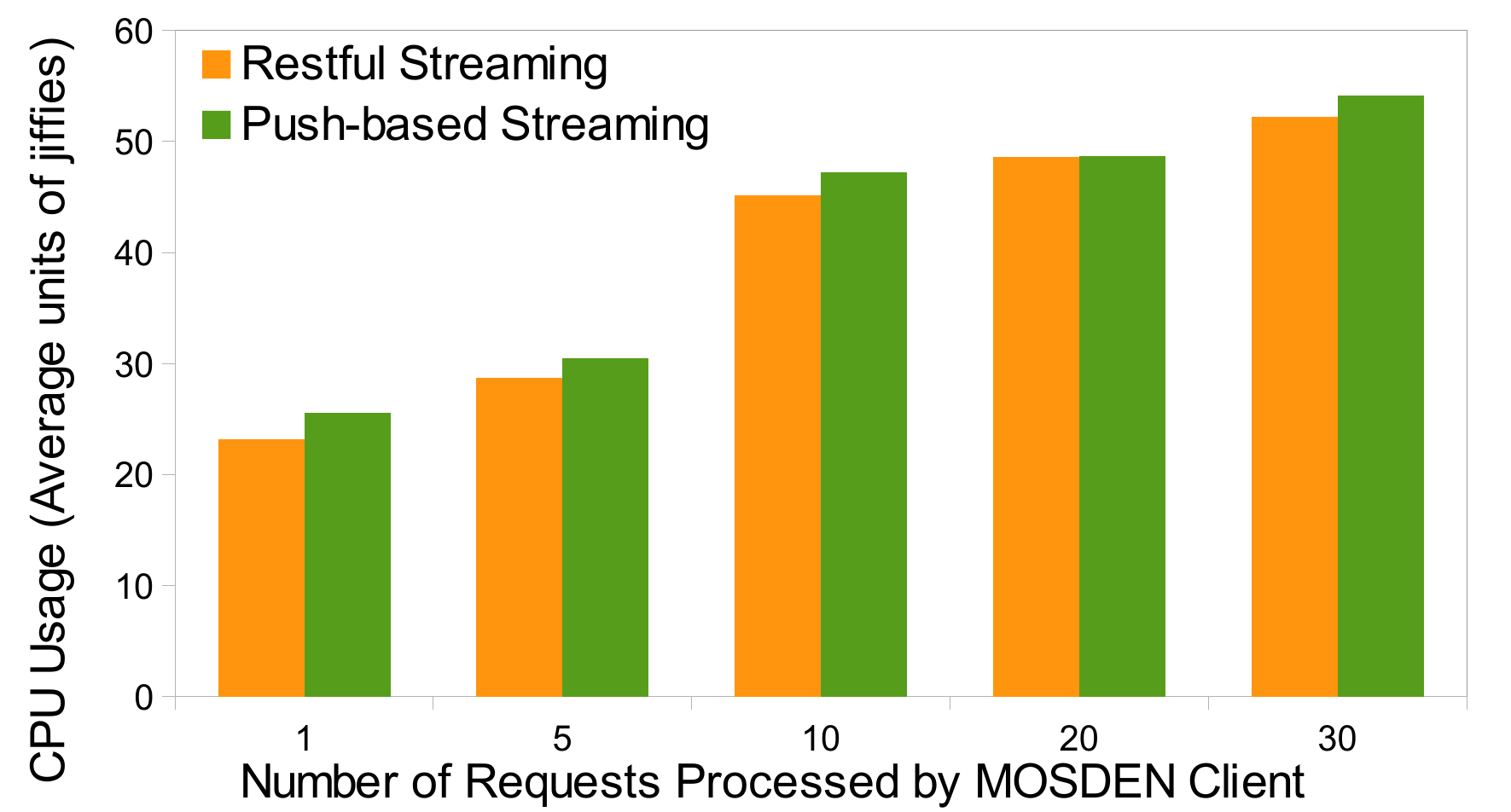}
\captionsetup{justification=centering, labelsep=period}	
 \caption{Comparison of CPU Usage by MOSDEN Client - Sensing + Sending}
 \label{Experiment1}		
\end{figure}

\begin{figure}[t!]
 \centering
 \includegraphics[scale=0.45]{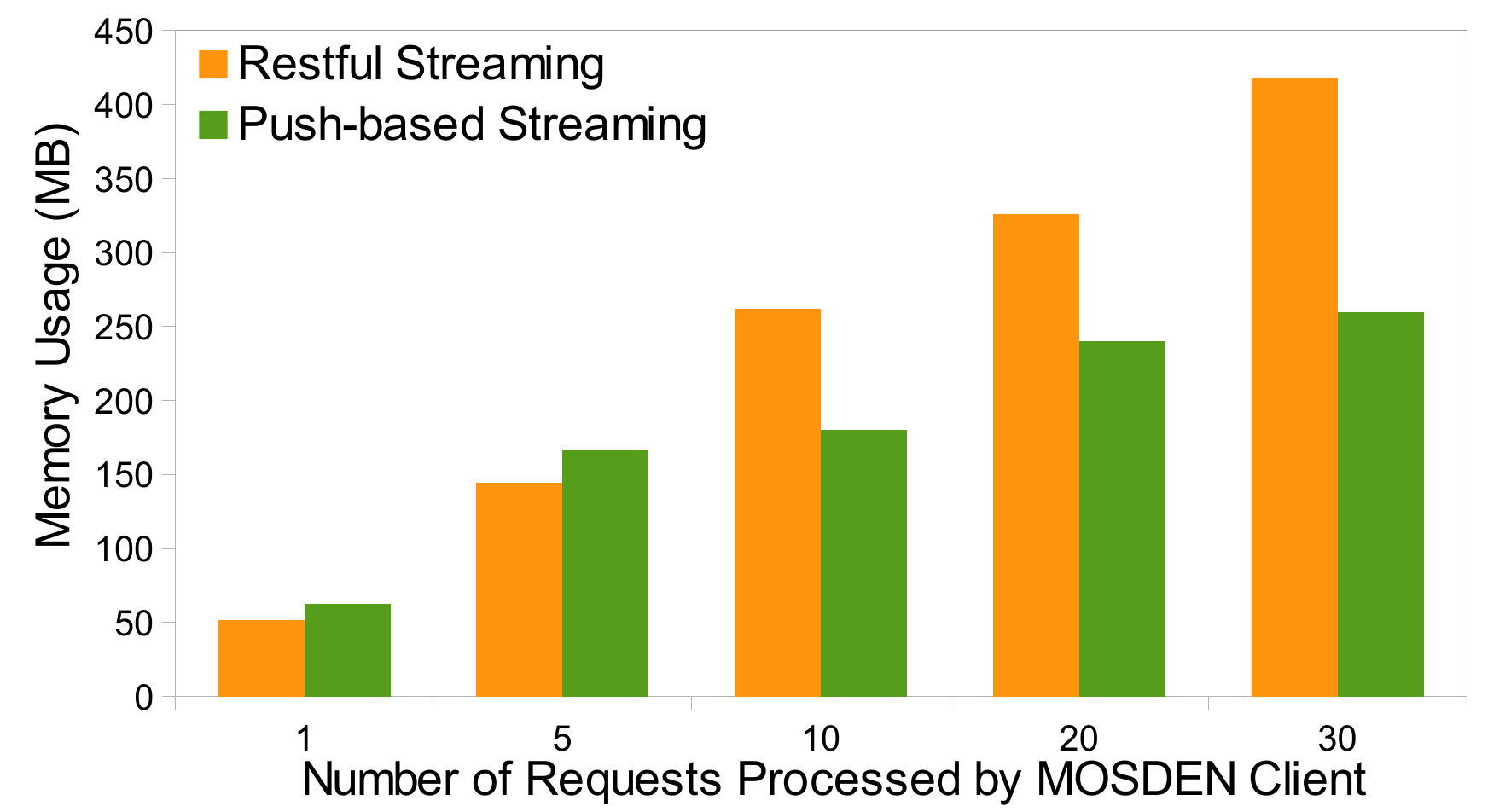}
\captionsetup{justification=centering, labelsep=period}	
 \caption{Comparison of Memory Usage by MOSDEN Client - Sensing + Sending}
 \label{Experiment2}		
\end{figure}

\begin{figure}[t!]
 \centering
 \includegraphics[scale=0.45]{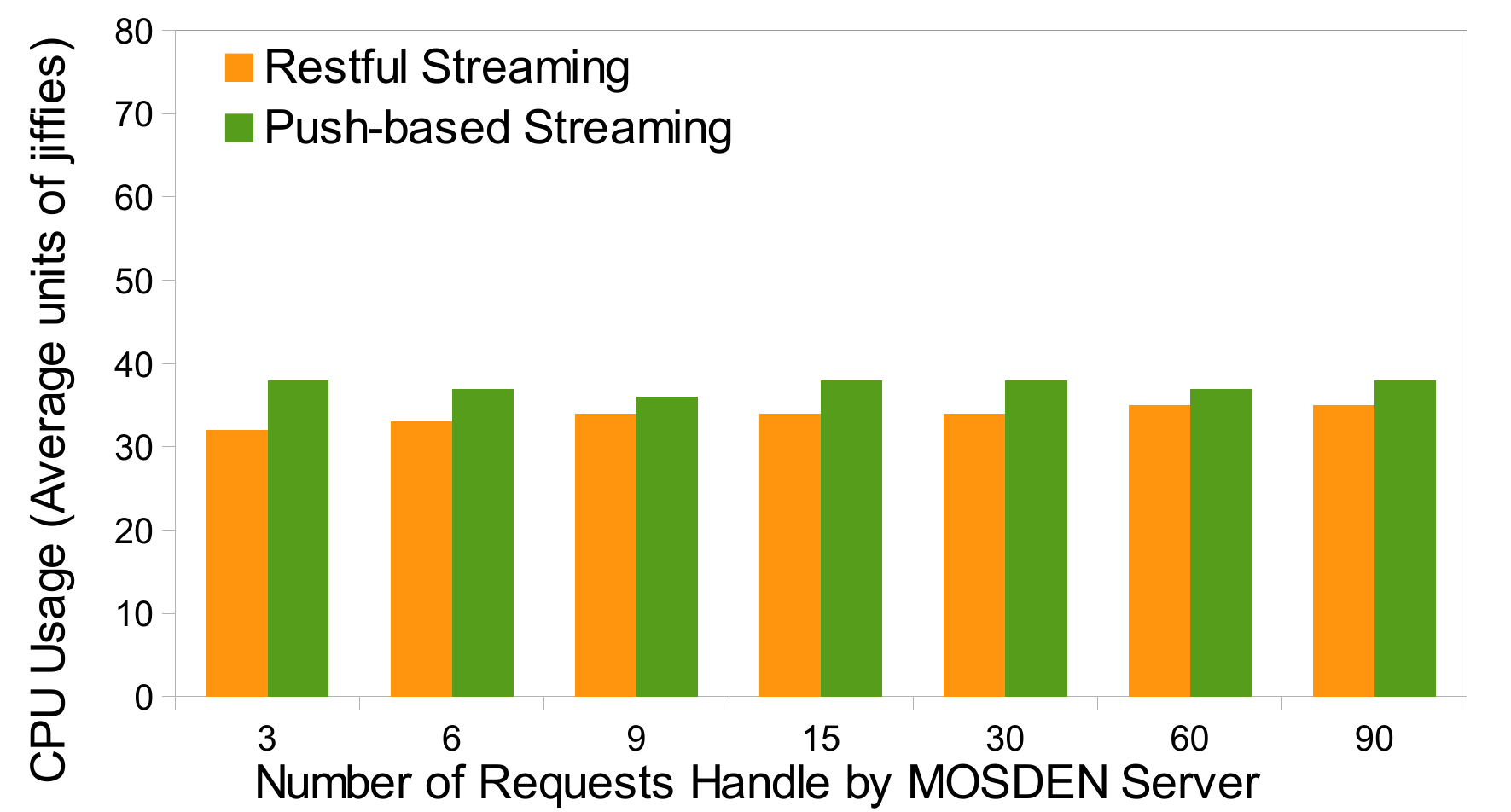}
\captionsetup{justification=centering, labelsep=period}
 \caption{Comparison of CPU Usage by MOSDEN Server}
 \label{Experiment5}		
\end{figure}

\begin{figure}[t!]
 \centering
 \includegraphics[scale=0.45]{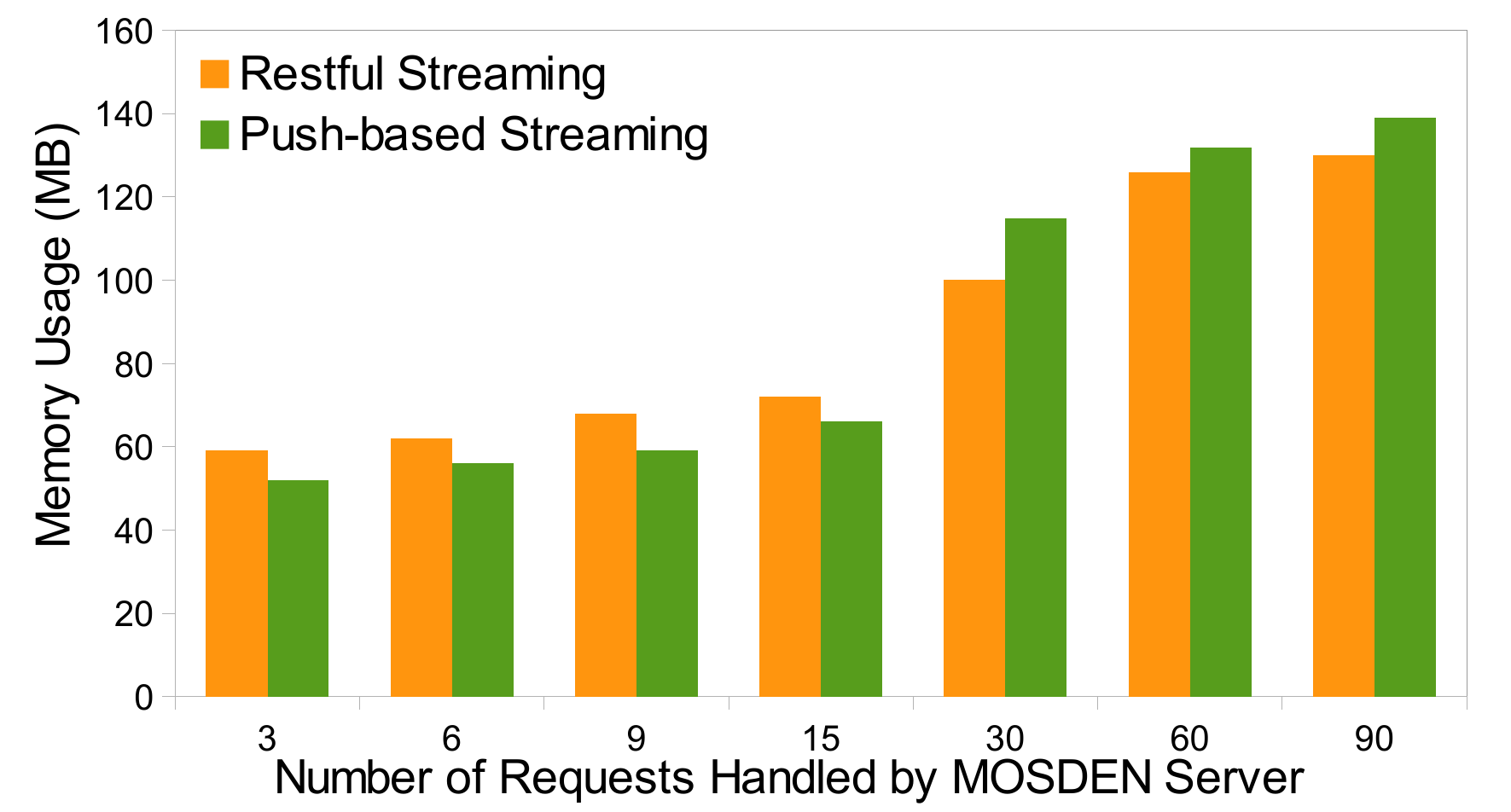}	
\captionsetup{justification=centering, labelsep=period}
 \caption{Comparison of Memory Usage by MOSDEN Server}
 \label{Experiment6}		
\end{figure}

For the query processing experiment, we used two data transmission methods between the MOSDEN client and server namely \textit{restful streaming} and \textit{push-based streaming}. Restful streaming is designed to have a persistent connection between the client and the server. On the other hand, the push-based approach makes a new connection every time to transmit data. Both these techniques can be used to perform communication between two (or more) distributed GSN or MOSDEN instances (i.e. GSN $\leftrightarrow$ GSN, MOSDEN $\leftrightarrow$ MOSDEN,  GSN $\leftrightarrow$ MOSDEN). The two approaches have their own strengths and weakness. The former is good for clients running MOSDEN that have a reliable data connection. The latter is useful for clients that need to work in off-line modes. The MOSDEN platform supports both the operations and the application developer has the choice to choose the best approach that satisfies the application requirements. The resource consumption experiment outcome of MOSDEN client is presented in Fig. \ref{Experiment1a}, \ref{Experiment2a}, \ref{Experiment1} and \ref{Experiment2}. The resource consumption experiment outcomes of MOSDEN server is presented in Fig. \ref{Experiment5} and \ref{Experiment6}.

Fig. \ref{Experiment1a} and \ref{Experiment2a} presents the CPU and memory consumption of MOSDEN client when performing sensing operation. We computed the memory and CPU consumption of the two devices independently due to the difference in their memory and processing capabilities. The memory allocation to MOSDEN is entirely managed by the Android operating system depending on available memory. As its can noted, MOSDEN has very little memory and CPU footprint for continuous operation even when the number of sensors connected increase to 13. This clearly validates the scalability of MOSDEN client to work under collaborative sensing application consuming significantly less resources. Fig. \ref{Experiment1} and \ref{Experiment2} illustrates the difference in CPU and memory usage of MOSDEN client during sensing and sending operations. The experiments observes the variation in CPU and memory consumption when the number of requests increases. According to Fig. \ref{Experiment1}, it is evident that restful streaming is marginally better than push-based streaming from CPU consumption perceptive.  On contrast, restful streaming consumes more memory than push-based streaming as depicted in Fig. \ref{Experiment2}. One reason for such a outcome could be attributed to the overheads involved in maintaining a persistent network connections. In both the cases, the MOSDEN client performed well to handle data capture, processing, storage and querying operations. Further, as it can be noted from the result in Fig. \ref{Experiment1}, the memory consumption increases to 400MB when MOSDEN is processing data concurrently from 30 sensors. The device used for this experiment was the Nexus 7 tablet (D2) with an available memory of 1 GB (average among most current day smartphones and tablets). As android manages memory allocation and the tablet was only running the device monitoring application, MOSDEN was allocated more memory as needed. When there is contention from other applications, the memory allocation to MOSDEN might decrease. Under such circumstances, MOSDEN will still perform significantly well which is further justified by the query response latency experiments presented later. Moreover, newer devices such as Google Nexus 5 and Samsung Galaxy S5 have over 3GB of on-board memory which will significantly increase the performance of MOSDEN. 

In Fig. \ref{Experiment5} and \ref{Experiment6}, we compare the performance of restful streaming and push-based streaming techniques in terms of CPU and memory usage by the mobile device (D1) functioning as MOSDEN Server (Fig. \ref{Figure:Setup}(b)). The experiments compute the difference in CPU and memory usage by MOSDEN server when the number of requests increases. According to Fig. \ref{Experiment5} restful streaming is better than push-based in terms of CPU usage while as indicated in Fig. \ref{Experiment6} push based streaming is slightly better that restful streaming in terms of memory consumption. This is due to the fact, push-based makes connection on-demand hence requiring more CPU and less memory while restful maintains a constant connection consuming less CPU and more memory (connection maintenance overhead). Further, it is important to note that both techniques maintain the same amount of CPU consumption over time despite the increase in requests it handles. Additionally, MOSDEN server consumes significantly less amount of memory in comparison to MOSDEN client. One reason is that MOSDEN client performs sensing, processing and local storage activities in addition to sending data to the server. In contrast, MOSDEN server performs query processing task only (from clients). As we mentioned earlier, when the number of requests handled by MOSDEN increases (give that no other tasks are performed), restful streaming technique performs better in term of both CPU consumption and memory consumption.

\subsubsection{Storage Requirements Experiment}
\begin{figure}[b!]
 \centering
 \includegraphics[scale=0.45]{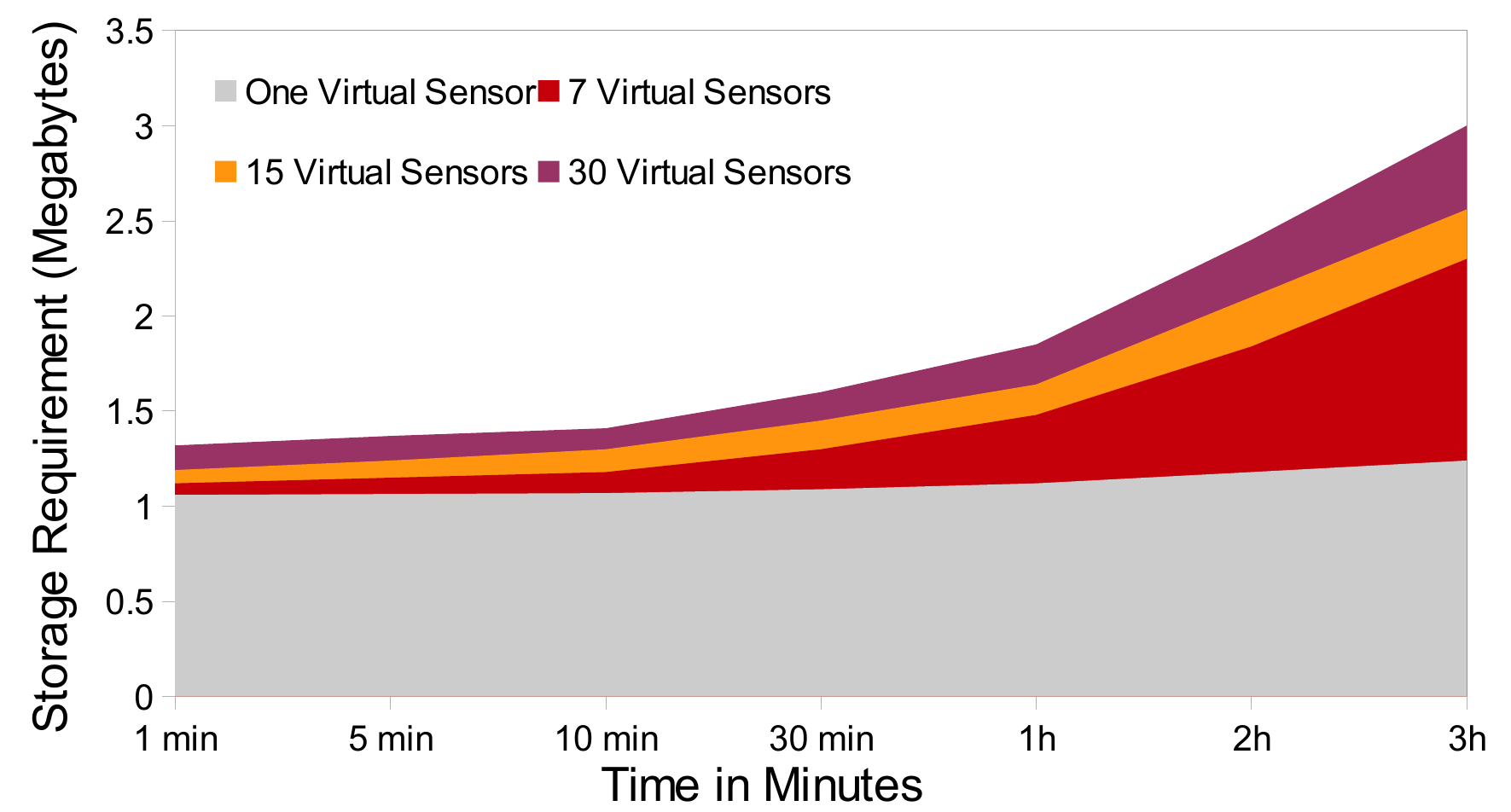}	
\captionsetup{justification=centering, labelsep=period}
 \caption{Storage Requirement of MOSDEN client}
 \label{Experiment4}		
\end{figure}

In this experiment we examine  how storage requirements vary when number of sensors handled by the MOSDEN client increases. For this experiment, we used Setup 1 in \ref{Figure:Setup}. All the sensors on-board the client mobile device (i.e. accelerometer, microphone, light, orientation, proximity, gyroscope, magnetic, pressure) are used as sensor sources. Sampling rate for sensors are configured as one second. The D1 (Setup 1) has been configured to receive data request from the server in an one second interval. The experiment was conducted for three hours. The exact storage requirements depend on multiple factors such as number of active sensors sending data, number of data items generated by the sensor\footnote{E.g. accelerometer generates 3 data items i.e. x, y, and z while temperature sensor generate one data item}, sampling rate, and history size \cite{P022}. We used external sensor to increase the number of sensors connected to MOSDEN during the experiment in order to examine the behaviour of MOSDEN from a storage requirement perceptive. The results of the experiment is presented in Fig. \ref{Experiment4}.

According to the outcome shown in Fig. \ref{Experiment4}, storage  requirements are linear i.e. the increase in storage changes at a constant rate depending on the history-size. History-size defines how much data record needs to be stored at a given time. Large history sizes can be used for summarising purposes or archival purposes. However, the amount of storage in easily predictable due to history size, because MOSDEN always deletes old items in order to accommodate new data items. In MOSDEN, storage can be easily controlled by changing the history-size. Specially, for real time reasoning history can be set to one. Considering all the above factor, it is fair to conclude that modern mobile devices have the storage capacity to store sensor data collected over long period of time. 
 
\subsubsection{Query Processing Experiment}
\begin{figure}[b!]
 \centering
 \includegraphics[scale=0.45]{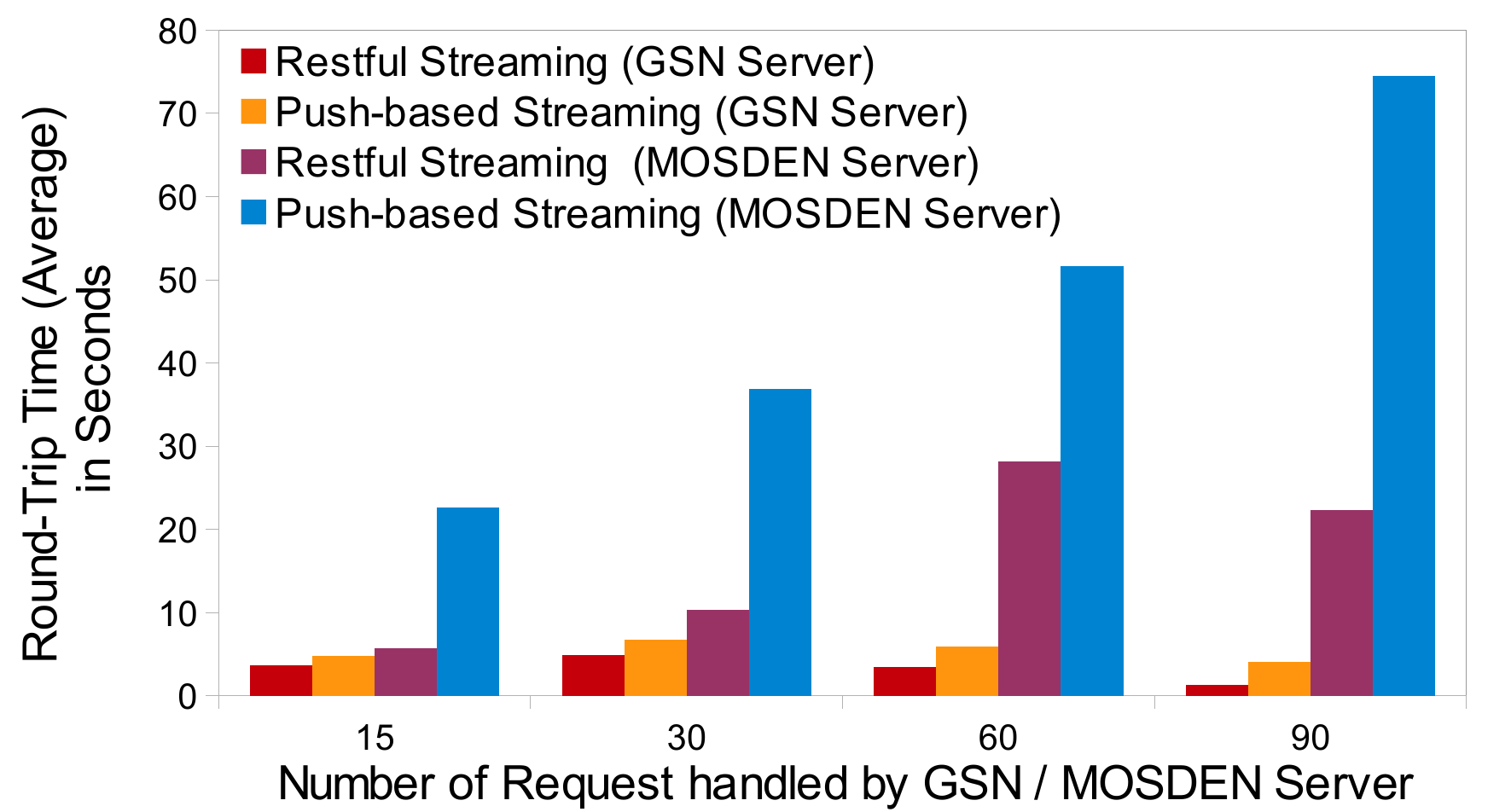}
\captionsetup{justification=centering, labelsep=period}	
 \caption{Comparison of Round-trip Times}
 \label{Experiment7}		
\end{figure}

\begin{figure}[b!]
 \centering
 \includegraphics[scale=0.45]{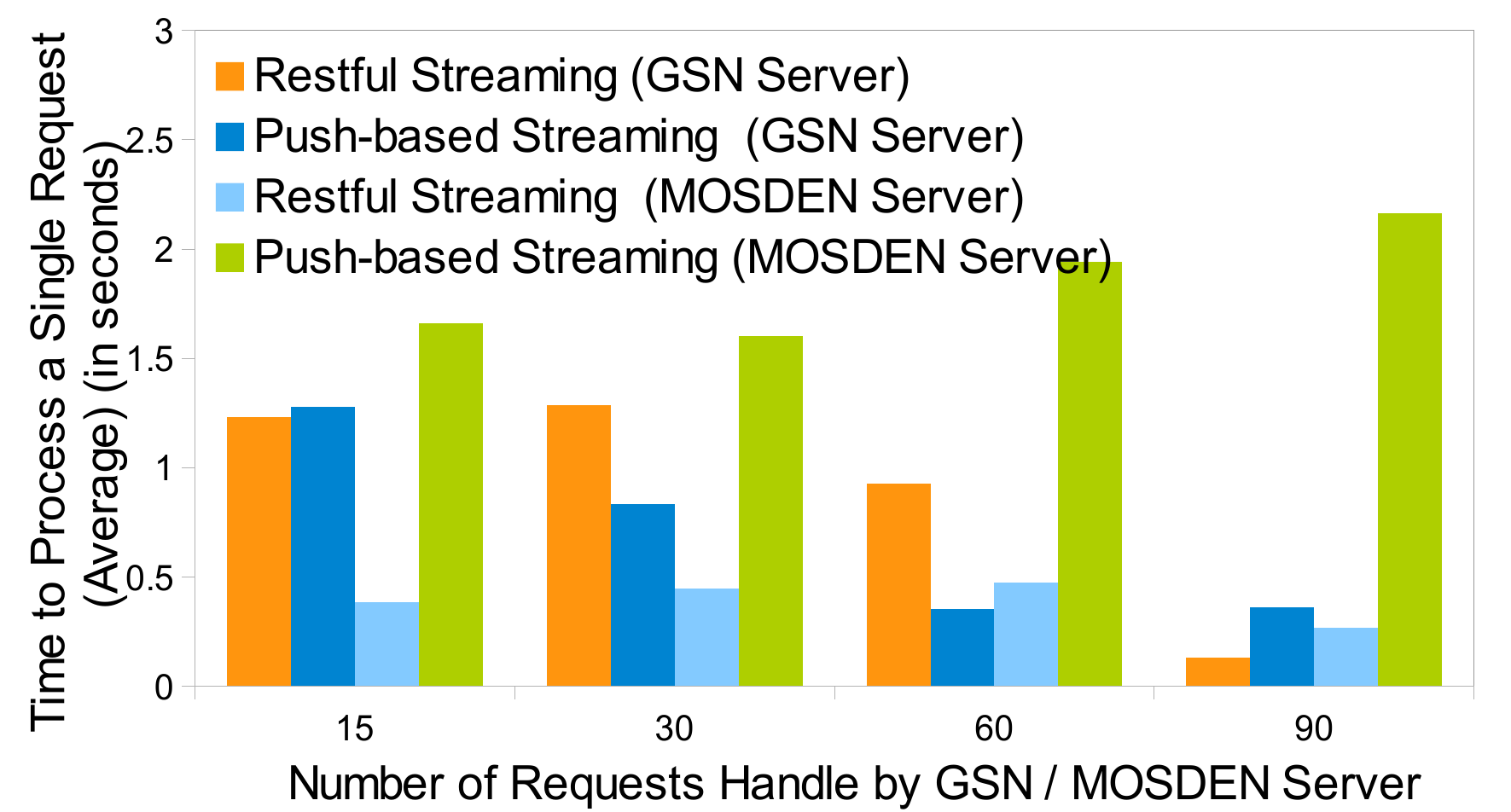}
\captionsetup{justification=centering, labelsep=period}	
 \caption{Comparison of Data Retrieval and Processing Ability}
 \label{Experiment8}	
\end{figure}

In this section, we present results of MOSDEN servers' query processing efficiency evaluation. To measure query performance, we evaluate how the round trip time\footnote{The round-trip time is the time taken for the server MOSDEN instance to request a data item from a given virtual sensor on a client MOSDEN instance. The total time is computed as the interval elapsed between server request and client response.} is impacted when the number of requests handled MOSDEN server (D1 in Fig. \ref{Figure:Setup}(b)) increases. Both restful streaming and push-based streaming techniques are evaluated separately. As a comparison, we compute the round-trip time in processing  a request by GSN server (D5 in Fig. \ref{Figure:Setup}(a)). Further, we also evaluate and compare the amount of time (average) it takes to process a single request\footnote{Time taken to process a single request is the time interval elapsed between two subsequent requests made by the server to any client irrespective of the virtual sensor}. This is different from round trip time and is calculated as denoted in Equation \ref{equation1}. The results of the experiments are presented in Fig. \ref{Experiment7} and \ref{Experiment8}.

\begin{equation}
	\begin{split}
	\textit{ Average time to process single request} \\
  = \frac{\textrm{Duration of the Experiment}}{\textrm{Total number of Round Trips Completed}}
 \label{equation1}
 \end{split}
\end{equation}
 
 \begin{figure*}[t!]
  \centering
  \includegraphics[scale=0.45]{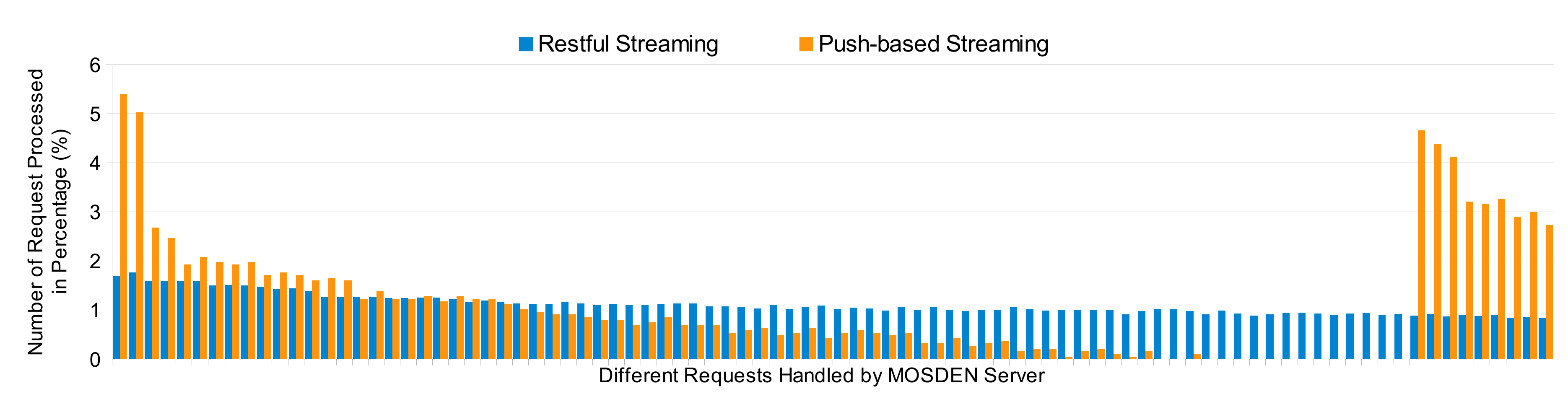}
 \captionsetup{justification=centering, labelsep=period}	
  \caption{Comparison of Requests Processing Variation}
  \label{Experiment9}		
 \end{figure*}
 
 \begin{figure*}[t!]
  \centering
  \includegraphics[scale=0.45]{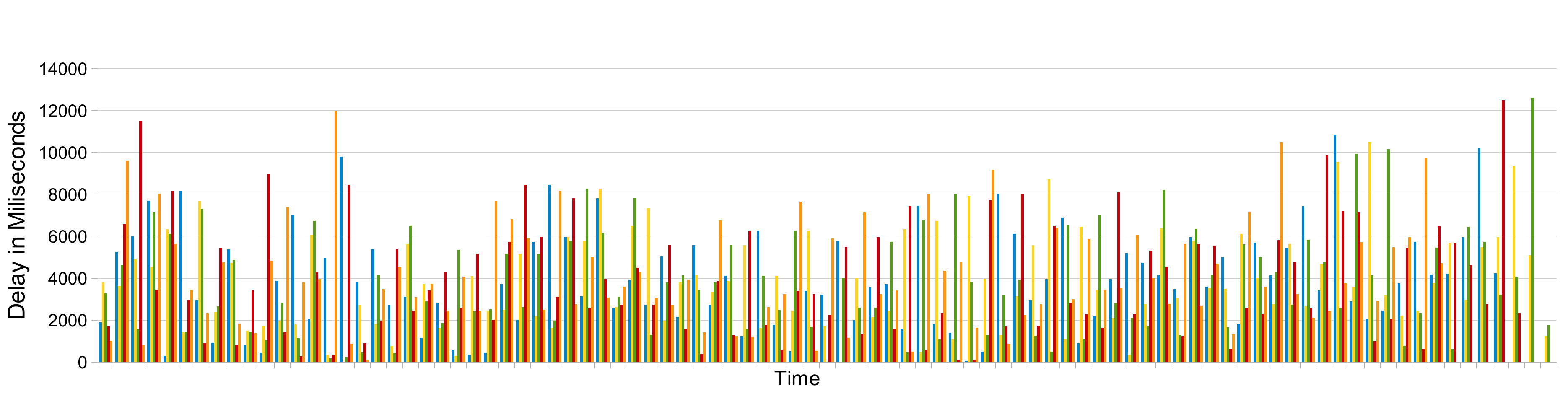}
 \captionsetup{justification=centering, labelsep=period}	
  \caption{Variation of round-trip time (delay / latency) over a period of time where seven requests are being processed}
  \label{Experiment10}		
 \end{figure*}

According to Fig. \ref{Experiment7}, it is clearly evident that resource constrained device such as mobile phones take more time to perform computations. As a result delay time is comparatively high when the server node is a mobile device in contrast to a computer-based processing node. Further it has been observed that (also we predicted in earlier section), push-based technique has much larger delay time due to additional overheads involved in connection setup and tear down For laptop-based server instances, the reason for having much less round trip time when handling 90 requests (3 clients * 30 queries each) is due to the availability of more computational resources. However, when resource constrained devices play the role of a server node, the CPU and memory resources are limited hence resulting in greater round trip times. Fig. \ref{Experiment8} also shows the impact of increased overheads when using a push-based streaming technique.  It is important to note that, even though, the average round trip time is  higher as observed in Fig. \ref{Experiment7} (e.g. 20 seconds when handling 90 requests) when restful steaming techniques is used, the amount of time taken to make subsequent requests by the server is mush less (e.g. less than a second when handling 90 requests) as observed in Fig. \ref{Experiment8}. This outcomes is further validated by results of the following experiment. 

In Fig. \ref{Experiment9} and \ref{Experiment10}, we presents results of the experiments (Fig. \ref{Figure:Setup}-Setup 2) that examine how each request was processed. We compared the performance using both restful streaming and push-based streaming. In this experiment, we configured MOSDEN server to make 30 requests to each of the three distributed client MOSDEN instances. We conducted the experiment for a fixed interval of time. Later, we calculated using Equation \ref{equation2}, the number of round-trips completed by each request and plotted them as a percentage. We denote the total number of round-trip requests completed for a virtual sensors $S$ as $S_{i}$ where $i$ is the virtual sensor identifier. The x-axis in Fig. \ref{Experiment9} represents $i$.
\begin{equation}
\begin{split}
\textit{Number of round-trips completed by each request} \\
  = \left ( \frac{\textrm{Number of Round trips Completed by } S_{i}}{\textrm{Total number of Round Trips Completed}  \sum_{i=1}^{n} S_{i}} \right ) \times 100
 \label{equation2}
 \end{split}
\end{equation}

 %Check for the experiment reference in the ()
According to Fig. \ref{Experiment9}, restful streaming technique allows each request to have fair amount of computational resources but push-based streaming does not. The main reason is attributed to the fact that restful streaming maintains a persistent connection between the client and server. When devices use push-based streaming, more computational resources are required to handle the connection setup and tear down Specially, when the number of requests that needs to be handled increases significantly, it places significant overheads on round-trip times for the push-based streaming approach as shown in Fig. \ref{Experiment9}. Due to restricted resources, under extremely high loads, in push-based streaming, there is a fair possibility that some requests made MOSDEN server to MOSDEN clients may not get executed at all. In Fig. \ref{Experiment10}, we visually illustrate how delay occurs in processing 90 requests (in Fig. \ref{Experiment9}, we only show 7 requests due to space limitation). Each request is shown in a different colour. Different requests (a combination of both restful and push based streaming queries were employed to compute the round-trip time) have different round-trip times depending on how processing capabilities and priorities of both server and client devices. This clearly shows the significance of the variation observed in previous experimental outcomes. Some requests (at some point of time) take only 6 milliseconds whereas some other requests take 12 seconds to complete a round trip.

\subsubsection{Energy Consumption Experiment}
\begin{figure}[b!]
   \centering
   \includegraphics[scale=0.45]{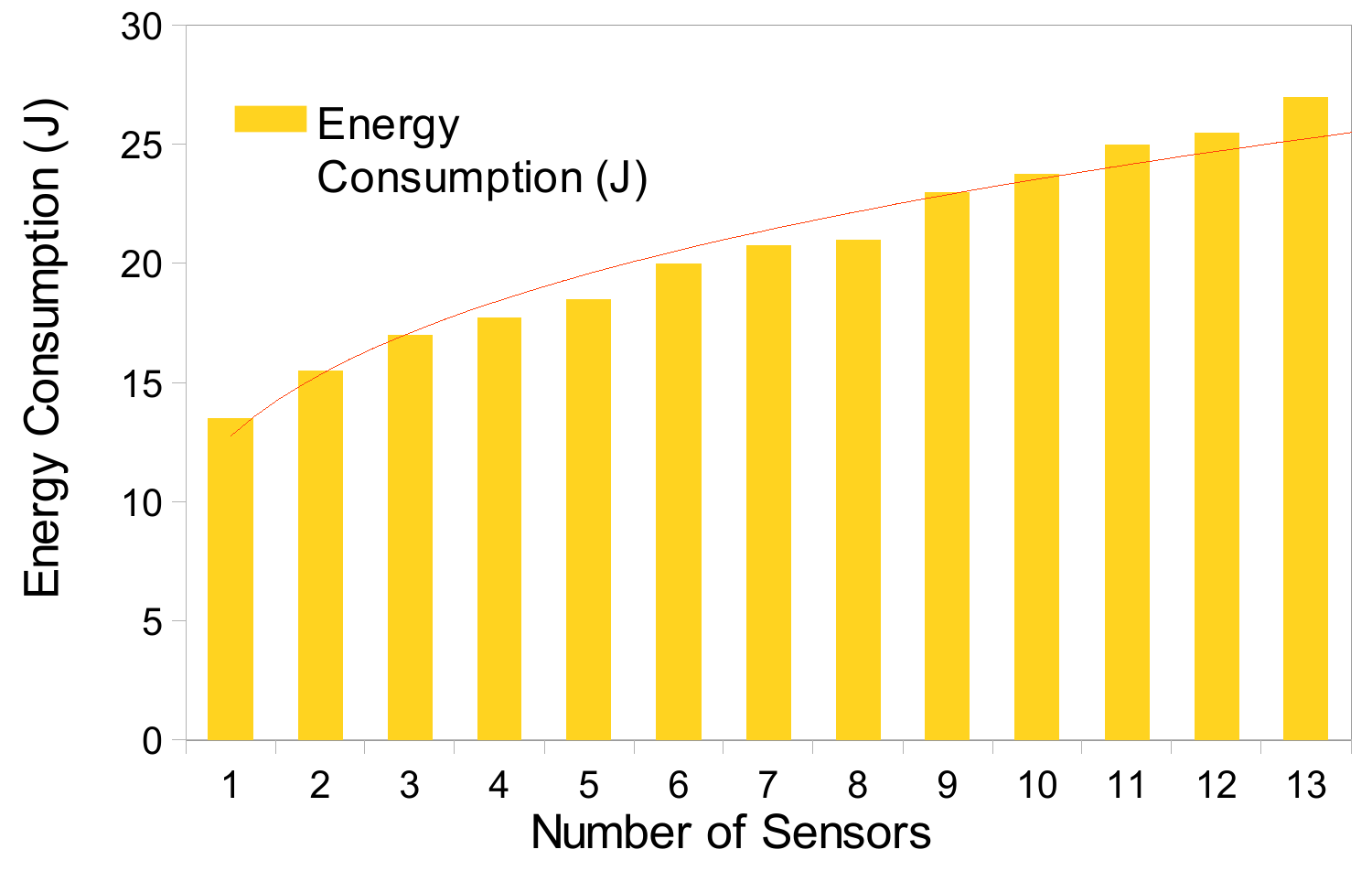}
  \captionsetup{justification=centering, labelsep=period}	
   \caption{Energy Consumption - MOSDEN Client (Sensing)}
   \label{Experiment11}		
  \end{figure}

\begin{figure}[b!]
     \centering
     \includegraphics[scale=0.45]{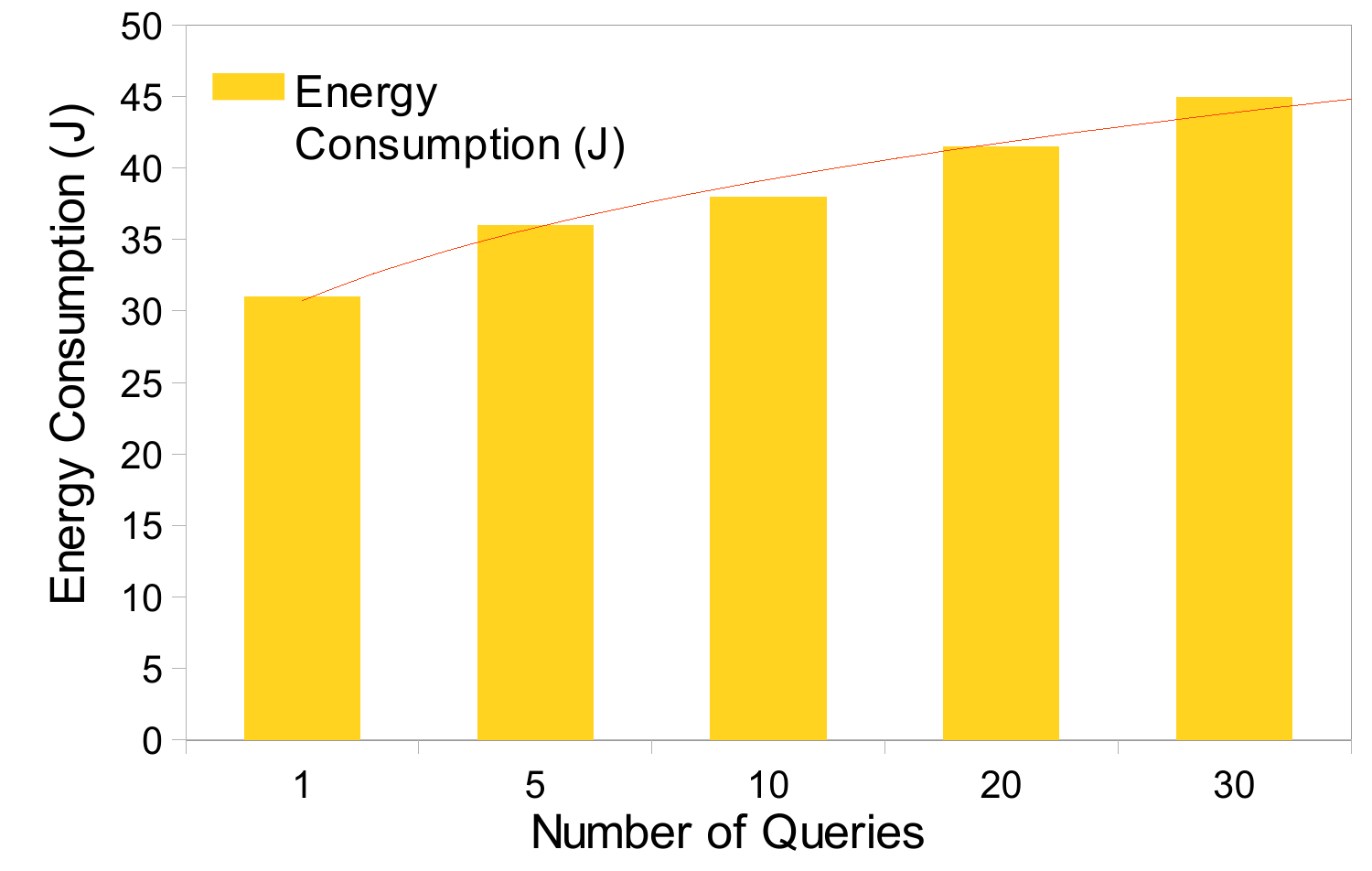}
    \captionsetup{justification=centering, labelsep=period}	
     \caption{Energy Consumption - MOSDEN Client (Sensing + Sending)}
     \label{Experiment11a}		
    \end{figure}

\begin{figure}[t!]
   \centering
   \includegraphics[scale=0.45]{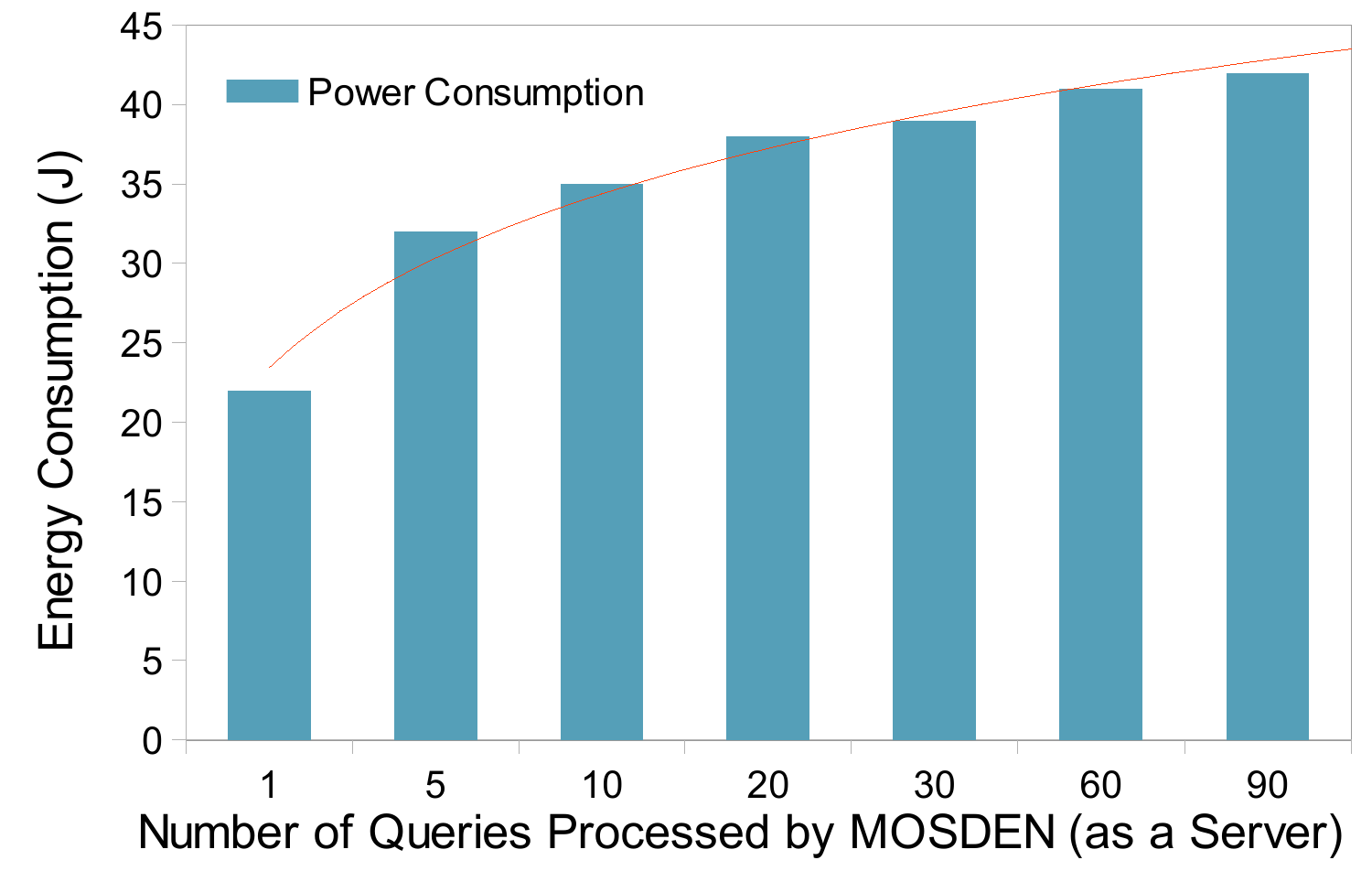}
  \captionsetup{justification=centering, labelsep=period}	
   \caption{Energy Consumption - MOSDEN Server}
   \label{Experiment12}		
  \end{figure}

This experiment evaluates the energy consumption of MOSDEN platform while functioning as both client and server. Energy consumption is vital consideration for any mobile device application. For this experiment, the MOSDEN client was tested with a 13 sensors including a combination of on-board sensors (accelerometer, gravity, gyroscope, linear acceleration, ambient temperature, light, pressure, relative humidity, magnetic field, orientation, proximity) and additional data source generators. We used the experimental setup depicted in Fig. \ref{Figure:Setup}(b). For the MOSDEN client, energy consumption for sensing only and sensing + sending operations were measured. The MOSDEN server was responsible to request data from 3 distributed clients and process the response instantaneously. Each MOSDEN client was issued 30 queries by the MOSDEN server. This resulted in MOSDEN server processing 90 queries in total (30 x 3). For this experiment, we chose the restful data streaming approach as a persistent data connection involves longer usage of Wi-Fi connection. During the experiment continuous requests with sampling rate of 1 second were made by MOSDEN server. The experimental outcomes are presented in Fig. \ref{Experiment11}, \ref{Experiment11a} and \ref{Experiment12}.

According to the results in  Fig. \ref{Experiment11}, \ref{Experiment11a} and \ref{Experiment12} it can be concluded that MOSDEN functions energy-efficiently under extreme loads (MOSDEN client sensing and processing 30 requests while MOSDEN server processing 90 requests). The experimental outcome clearly validates and supports this inference. We note, the average energy consumption by MOSDEN client over a 30 minute time window for 13 virtual sensors and MOSDEN server processing 90 requests was $\approx$ 40J. It is to be noted, in our energy consumption experiment we did not consider LCD consumption as this is entirely dependent on user's usage pattern. Further, we controlled the amount of data transmitted during experimentation by increasing and decreasing the number of queries sent to MOSDEN instance. Changes to the size of sensed data did not impact the energy consumption significantly.

\subsubsection{Discussion}Overall MOSDEN performs extremely well in both server and client roles in collaborative environments. MOSDEN (as a server) was able to handle 90 requests (i.e. 180 sub requests including 90 requests/90 responses) where each request has a sampling rate of one second. This resulted in a MOSDEN server (running on a mobile device) processing 5400 data points (90 requests * 60 seconds) every minute from distributed clients. Similarly, a MOSDEN client was stress tested with up to 13 virtual sensors which included a combination of on-board sensors and additional data source generators. Hence, the MOSDEN clients was processing 1800 data points (30 queries on 13 sensors * 60 seconds) every minute. It is to be noted, that for evaluation purposes and to test the energy efficiency, resource consumption, performance and scalability of MOSDEN, we conducted experiments on MOSDEN server and client under extreme loads. Such processing is intensive and rare in real-world applications. However, our experiments showed that MOSDEN can withstand such intensive loads proving to be a scalable, performance oriented and energy efficient platform for deploying large-scale \osensing applications. Under such extensive loads, considering the battery rating of Google Nexus 7 (16Wh), the MOSDEN server and MOSDEN client (sensing + sending) in continuous processing mode with 1 second sampling rate can last $\approx$ 20 hours while the MOSDEN client in sensing only mode can last $\approx$ 35 hours. If MOSDEN is configured to collect data from 10 different sensors and handle 30 requests (typical of real-world situations), it can perform real-time sensing with delay of 0.4 - 1.5 seconds. 

%When the server node is a computer (D5 as explained in Section \ref{sec:E:Experimentation Testded}) both restful streaming and push-based streaming work extremely well without visible significant differences. However, when the server node is a mobile device, which runs MOSDEN, restful streaming performs approximately 6 times better than push-based technique. 

\section{Conclusion and Future Work}
A mobile \osensing application development framework must scale from an individual user to user communities (tens of thousands of users). In this paper, we proposed MOSDEN, a collaborative mobile platform to develop and deploy \osensing applications. MOSDEN differs from existing \osensing platforms by separating the sensing, collection and storage from application specific processing. This unique feature of MOSDEN renders it an easy-to-use, reusable framework for developing novel \osensing applications. We proposed the architecture of the MOSDEN framework. We then demonstrated its ease of use and minimal development effort requirement by developing a proof-of-concept noise pollution application. We validated MOSDEN's performance, energy efficiency, resources consumption and scalability when working in distributed collaborative environments by extensive evaluations under extreme loads resolving and answering queries from external sources (MOSDEN instances and GSN in the cloud). Overall MOSDEN is extremely energy and resource efficient and performs exceedingly well under high degrees of load in collaborative environments validating its suitability to develop large-scale \osensing applications. 

Our next step is to explore protocols for dynamic discovery, load balancing and task allocation among MOSDEN sensing and processing resources in a typical mobile ad-hoc network scenario. The aim of the extension is to dynamically distribute a collective task to a set of MOSDEN clients and servers autonomously to achieve a common goal.

\acks Part of this work has been carried out in the scope of the ICT OpenIoT Project which is co-funded by the European Commission under seventh framework program, contract number FP7-ICT-2011-7-287305-OpenIoT. The authors acknowledge help and support from CSIRO Sensors and Sensor Networks Transformational Capability Platform (SSN TCP). 
\bibliography{Bibliography}
\bibliographystyle{icstnum}

\end{document}